\documentclass[preprint,12pt]{elsarticle}

\usepackage{amssymb}
\usepackage{epsfig}
\usepackage{amssymb}
\usepackage{bm}
\usepackage{braket}
\usepackage{here}
\usepackage{color}
\usepackage{graphicx}
\usepackage{epstopdf}
\usepackage{float}
\usepackage{color}
\usepackage{siunitx}
\usepackage{soul,color}
\usepackage[colorlinks=true,linkcolor=green,citecolor=blue,linkbordercolor={1 0 0}]{hyperref}
\usepackage{setspace}
\soulregister\cite7
\soulregister\ref7
\usepackage{ulem}
\usepackage{caption}
\usepackage{hyperref}
\usepackage{array,multirow,makecell} 
\usepackage{caption,tabularx,booktabs}
\usepackage{keyval}
\usepackage{graphicx}
\usepackage{comment}

\journal{Nuclear Physics A}

\begin{document}

\begin{frontmatter}



\title{Investigation of giant dipole resonance  in Mo isotopes within TDHF theory}


\author{A. Ait Ben Mennana $^{a}$, M. Oulne $^{a}$}

\affiliation{organization={ High Energy Physics and Astrophysics Laboratory, Department of Physics,
		Faculty of Sciences SEMLALIA},
            addressline={}, 
            city={Marrakesh},
            postcode={P.O.B. 2390}, 
            country={Morocco}}

\begin{abstract}
The isovector giant dipole resonance (IVGDR) in the chain of even-even Mo isotopes is investigated within the time-dependent Hartree-Fock (TDHF) using the Skyrme force Sly6. The GDR calculated in  $ ^{92-108}\text{Mo}$ are presented, and compared with the  available experimental data. An overall agreement  between them is obtained. Moreover, the dipole strength in  $ ^{102-108}\text{Mo}$ is predicted. Shape phase transition from spherical to oblate as well as shape coexistence (A $\sim$ 100) in Mo isotopes are also investigated in this work. In addition, the correlation between the deformation splitting $\Delta E$ and the quadrupole deformation parameter $\beta_{2}$ is studied. The results confirm that $\Delta E$ is proportional to the deformation of nucleus. We also discuss the dependence of GDR strength on some nuclear properties of Skyrme forces, such as the asymmetry energy $a_{s}$. We find that $a_{s} = 32 MeV$, corresponding to SLy6, is quite consistent with experimental data. 

\end{abstract}


\end{frontmatter}

\section{Introduction}\label{sec1}
\qquad Collective excitations in quantum many-body systems such as atomic nuclei are a common phenomenon. An example of theses excitations is nuclear giant resonances (GRs) which are considered as a collective motion of many, if not all, particles in the nucleus \cite{harakeh2001}. Among all the possible
modes of collective excitations, the isovector giant dipole resonance (IVGDR) is the first collective excitation observed in nuclei \cite{baldwin1947}. It has been explained as a collective vibration mode of neutrons oscillating against protons in the nucleus \cite{goldhaber1948}. The GDR is a most pronounced characteristic of the excitation spectrum of all nuclei (except deuterons) in the nuclide chart, giving crucial clues for understanding nuclear structure and collective dynamics. 
It has been widely studied in spherical, transitional and deformed nuclei with many experimental investigations \cite{veyssiere1973,gurevich1976,carlos1971,carlos1974,berman1975,ceruti2017} and theoretical methods \cite{maruhn2005,reinhard2008,colo2008,yoshida2011,benmenana2020}.\\

The evolution of the GDR properties (centroid energy E, shape, width $\Gamma$...) as a function of the mass number A of nuclei has been extensively studied in many experimental and theoretical works ( see for example Refs. \cite{carlos1971,beil1974,reinhard2008,wang2017}). The centroid energy E been inversely proportional to A, was approximated as $ E \simeq 80 A^{-1/3}$ Mev for spherical nuclei \cite{ring1980}. Therefore, the centroid energy of GDR can provide direct information about nuclear size. On the other hand, the width $\Gamma$ represents an important characteristic of any giant resonance, as it provides valuable information about the excitation and decay of giant resonance. Based on model, it is often expressed by the sum of three terms \cite{harakeh2001}:
$ \Gamma = \Delta \Gamma + \Gamma^{\uparrow} + \Gamma^{\downarrow}$, where $ \Delta \Gamma$, called Landau damping, arises from fragmentation of the initial dipole excitation  into one particle-one hole(1p-1h) states, $ \Gamma^{\uparrow} $, called escape width, results from the coupling of the correlated 1p-1h state to the continuum, and $ \Gamma^{\downarrow} $ represents the spreading width arising from the coupling with 2p-2h, 3p-3h,..., np-nh configurations. The correlation between the width $\Gamma$ of GDR and the deformation parameter $\beta_{2}$ can be used as  a direct experimental probe to measure the nuclear deformation at finite temperature and angular momentum over the entire mass region \cite{mattiuzzi1997,pandit2013}. The GDR strength can provide an idea about the shape of nucleus. It has a single peak for heavier spherical nuclei, while it is more fragmented in light ones due to configuration splitting \cite{eramzhyan1986,harakeh2001}. For axially symmetric deformed nuclei, the GDR strength splits into two components, one corresponding to an oscillation along the symmetry axis (the K = 0 mode) and one to an oscillation perpendicular to it (the $\mid$K$\mid$= 1 mode) \cite{van1987,harakeh2001,speth1981}. The strength ratio for the two components provides the sense of the deformation. For a prolate nucleus, 2:1 in favor of the high-energy resonance component, and 1:2 for an oblate nucleus.\\

The deformed nuclei are of special interest in the GDR because the statically deformed ground-state gives rise to a very splitting of GDR in these nuclei. This correlation between nuclear deformation and splitting was first predicted theoretically by Okamoto \cite{okamoto1958} and Danos \cite{danos1958}, and then detected experimentally by Fuller and Weiss \cite{fuller1958}. The GDR in deformed nuclei has been studied extensively using different microscopic theoretical approaches based on Skyrme forces \cite{skyrme1956}, such as time-dependent Hartree-Fock (TDHF) \cite{mennana2021,maruhn2005}, Quasi-particle Random Phase Approximation (QRPA) \cite{yoshida2011}, and Separable Random-Phase-Approximation (SRPA) \cite{nesterenko2007}. Experimentally, the GDR  is induced by various ways such as photo-absorption~\cite{carlos1974, Masur2006} inelastic scattering \cite{ramakrishnan1996}, and $\gamma$-decay~\cite{gundlach1990}. The time-dependent Hartree-Fock (TDHF) \cite{dirac1930} has become, in recent years, a sufficiently mature technique to be used in realistic situations describing a range of nuclear dynamics. In particular, it was applied to giant resonance, giving a satisfactory results (see for example Refs. \cite{benmenana2020,maruhn2005,fracasso2012}). \\

In our previous studies~\cite{mennana2021}, we have studied the GDR in Sm isotopes with the Sky3d code \cite{sky3d} based on TDHF method using four Skyrme forces. All these forces  gave an accurate description of the GDR in spherical and deformed nuclei with a slightly advantage for the parametrization SLy6 \cite{CHABANAT1998}. Many previous works have studied the GDR in Molybdenum (Mo) isotopes, where Z=42. From the experimental point of view one can see for example Refs.\cite{beil1974,rusev2008,wagner2007}) and from the theoretical one Ref.\cite{ishkhanov2014}. Beil et al.\cite{beil1974} are studied the GDR in five stable even-even $^{92-100}\text{Mo}$ isotopes using a variable monochromatic photon beam. they observed a broadening of GDR as the number mass A increases.

The present work is aimed to study, some properties of the GDR in even-even $ ^{92-108}\text{Mo}$ nuclei. We extended this study to $^{108}\text{Mo}$ nucleus, i.e, nine Mo isotopes are considered in this work. This study is done with 
a fully-fledged TDHF approximation based on Skyrme effective interactions \cite{skyrme1956}, without any symmetry restrictions. Due to the open-shell nature of these nuclei, one should take account of pairing and deformation properties in this study. In the static step, the ground states of these isotopes are calculated with the static Hartree-Fock (HF) \cite{hartree1928} plus Bardeen-cooper-Schrieffer (BCS) approach \cite{bardeen1957}, where pairing is treated properly. In the dynamic step,  the strength of GDR is calculated by boosting the ground state with a dipole excitation.\\

The structure of this paper is as follows. Section \ref{sec2} gives a brief summary of nuclear giant dipole resonance (GDR) in deformed nuclei. Section \ref{sec3} describes briefly the TDHF approximation, and details of the numerical calculations. Our results and discussion are presented in Section \ref{sec4}. Finally, a summary is given in Section \ref{sec5}.
\section{Giant dipole resonance in deformed nuclei}\label{sec2}

\qquad For the ground state GDR, it is well known that for a deformed nucleus the resonance splits into several components. The centroid energy (frequency) of each component is inversely proportional to the radius $ R_{i}$ of the axis along which the vibration occurs \cite{speth1981}.
\begin{equation}{\label{E_R-1}}
	E_{i} = \hbar \omega \sim R_{i}^{-1} \sim A^{-1/3},
\end{equation} 
since $R$ is proportional to $ A^{1/3} $, where A is the mass number of the nucleus. This splitting has been observed experimentally \cite{carlos1974,berman1975} and treated theoretically by different models \cite{maruhn2005,reinhard2008,yoshida2011}.

The GDR spectra depend on the shape of the nucleus. Thus, for deformed prolate nuclei, the dipole strength splits into two components where one is associated to oscillations of neutrons against protons along the symmetry axis, and another, at higher energy and carrying twice the strength of the first, associated to oscillations perpendicular to the symmetry axis. The opposite situation occurs in the case of oblate nuclei. For triaxial nuclei, the GDR splits into three components with different oscillation,i.e.,  $\omega_{x} \neq  \omega_{y} \neq \omega_{z} $. In spherical nuclei, the GDR has a single peak especially for medium and heavy ones, which represents a superposition of three resonances with the same frequencies $\omega_{x} =  \omega_{y} = \omega_{z} $.

The energy splitting $\Delta E$ is related to the nuclear deformation, in which the   size of this splitting is proportional to the quadrupole deformation parameter $\beta_{2}$ \cite{okamoto1958,danos1958}:
\begin{equation}{\label{dE&beta}}
	\Delta E \propto \beta_{2},
\end{equation}
where $\beta_{2}$  defined as \cite{sky3d}

\begin{equation}{\label{beta}}
	\beta_{2} = \sqrt{a_{0}^2 + 2a_{2}^2}, \qquad   \text{with} \qquad a_{m} = \frac{4\pi}{5}\frac{Q_{2m}}{AR^2}. 
\end{equation}
where 
\begin{equation}{\label{Q_2m}}
	Q_{2m} = \int d\vec{r} \rho(\vec{r}) r^{2} Y_{2m}
\end{equation}

are the quadrupole moments, and $ Y_{2m} $ are the spherical harmonics. On the other hand, there is a correlation between the GDR width $\Gamma$ and $\beta_{2}$, especially for medium and heavy nuclei \cite{ishkhanov2011}. They are minimal for magic nuclei, and simultaneously increase as soon as the number of neutrons increases \cite{carlos1971,mennana2021}. For light nuclei ($ A<60 $), the nuclear deformation is not the only reason behind GDR broadening. Many factors can affect the shape of the GDR, such as configuration and isospin splitting\cite{eramzhyan1986,ishkhanov2015}. For more details see for example Ref.\cite{ishkhanov2021}
\section{Theoretical framework and numerical details}\label{sec3}
\subsection{Time-dependent Hartree-Fock method}
\qquad The time-dependent Hartree-Fock (TDHF) is a self-consistent mean field theory which was originally proposed by Dirac in 1930 \cite{dirac1930}. It has been extensively applied in nuclear physics calculations like heavy-ion reactions\cite{maruhn2006,simenel2018} and nuclear giant resonances\cite{blocki1979,stevenson2004}. A detailed discussion of the TDHF approximation can be found in many papers (see for example Refs.\cite{bonche1976,kerman1976,koonin1977}). Here, a brief
introduction of the TDHF theory is presented as follows.\\
The TDHF equations are obtained from the variation of the action S defined as:
\begin{eqnarray}{\label{action_S}}
	S & = & \int_{t_1}^{t_2} dt \bigg( \bra{\Psi(t)}\big(i\hbar \frac{\partial }{\partial t} - \hat{H} \big )\ket{\Psi(t)}\bigg) \nonumber \\
	& = & \int_{t_1}^{t_2} dt \bigg( i\hbar \sum_{i=1}^{N} \bra{\psi_{i}}\frac{\partial }{\partial t}\ket{\psi_{i}} - E[\psi_{i}]\bigg) 
\end{eqnarray}
with respect to the wave functions $ \psi_{i}^* $, where \textit{$ \ket{\psi_{i}} $} are the occupied single-particle states, $ t_{1} $ and $ t_{2} $ define the time interval, where the action S is stationary between them, and $ E = \bra{\Psi} \hat{H} \ket{\Psi} $ is the total energy of the system.\\
The variation of Eq. (\ref{action_S}) leads to the coupled TDHF equations \cite{flocard1978}
\begin{equation}{\label{TDHF}}
	i\hbar\frac{\partial \psi_{i}(t)}{\partial t} = \hat{h}[\rho(t)]\psi_{i}(t), \qquad  i = 1,.....,A,
\end{equation}
where $ \hat{h} $ is the single-particle Hartree-Fock Hamiltonian which depends on the nucleon density $\rho$, A is the total number of nucleons.\\
Since these equations (\ref{TDHF}) are nonlinear, they are solved \textit{iteratively} in time with
a small time step increment $ \Delta t $. Our calculations were performed
with $\Delta t$ = 0.2 fm/c $\simeq$ $6 \times 10^{-25} s$. Over the small intervals [t, t + $\Delta t$], the HF Hamiltonian is assumed to be constant. To conserve the total energy E, it is necessary to apply a symmetric algorithm by time reversal, and therefore to estimate the Hamiltonian at time $ t + \frac{\Delta t}{2} $ to evolve the system between time $ t \;\text{and}\; t+ \Delta t$ \cite{bonche1976}
\begin{equation}{\label{evol_psi}}
	\ket{\psi(t+\Delta t)} \simeq e^{-i\frac{\Delta t}{\hbar}\hat{h}(t+\frac{\Delta t}{2})}\ket{\psi(t)}.
\end{equation} 
More details regarding numerical procedures for solving the TDHF equations can be found in Refs.\cite{bonche1976,flocard1978,simenel2012,koonin1977}.

\subsection{Details of Calculations}
\qquad In this work, all calculations were performed with the Sky3D code\cite{sky3d}, which solves the static HF as well as the TDHF equations. These calculations are performed on the three-dimensional (3D) Cartesian coordinate grid space, with no symmetry restrictions and the full Skyrme energy functional is used including
the spin-orbit and most important time-odd terms. The Skyrme force SLy6\cite{CHABANAT1998} was used in this study, which was found to give a satisfying description of the GDR for medium and heavy nuclei\cite{nesterenko2006,nesterenko2007,benmenana2020}.

In order to perform a TDHF calculation, an initial state is required. This state is obtained by iteratively solving the static HF + BCS equations 

\begin{equation}{\label{hf+bcs}}
	\hat{h}\psi_{i}(r)= \epsilon_{i} \psi_{i}(r), \qquad  i=1,....,\Omega,
\end{equation}
where $ \hat{h} $ is the single-particle Hamiltonian, $ \epsilon_{i} $ is the single-particle energy of the state  $ \psi_{i}(\bm{r})$ with the \textbf{r} coordinate is chosen to represent all the spatial, spin and isospin coordinates for brevity, and $\Omega$ denotes the size of the pairing-active space. In this work, we took the number of single-nucleon wave functions from the approximate relation \cite{sky3d,bender2003}: 
\begin{equation}{\label{N_max}}
	N_{q} + \frac{5}{3} N_{q}^{2/3},
\end{equation}
 where $ N_{q} $ refers to the number of protons (Z) or neutrons (N) in the nucleus under study. 
The static calculations are stopped  when sufficient convergence is achieved, i.e., we have reached minimum energy which represents a local minimum (isomeric state). In order to find global minimum, we redo the calculation several times with different initial configurations.	The state with lowest energy of all these minima is	
the global minimum.
We consider as convergence criterion the average single particle energy fluctuation given
by the expression \cite{sky3d}
\begin{equation}{\label{fluct_energy}}
	\sqrt{\bra{\psi} \hat{h}^{2}\ket{\psi} - \bra{\psi}\hat{h}\ket{\psi}^{2}},
\end{equation}

we took $10^{- 5}$ as a  convergence value for isotopes under study in this work. We used grids with 24 $\times$ 24 $\times$ 24 points
and a grid spacing of $ \Delta x = \Delta y = \Delta z$ = 1 fm in the x, y and z directions. The pairing is treated in the static calculation, it is frozen in the dynamic calculation i.e, the BCS occupation numbers are frozen at their initial values during time evolution.\\

In dynamic calculations, The nucleus is excited by multiplying the ground-state wave functions obtained from the HF  calculation by a an instantaneous initial dipole boost operator applied in the same manner to each single-nucleon state, i.e.,  
\begin{equation}{\label{boost}}
	\psi_{i}(r) \longrightarrow \psi_{i}(r,t=0)= \text {e}^{ib\hat{D}} \psi_{i}(r),
\end{equation}
where $\psi_{i}(r) $ represents the stationary wave function before boost,
b is the boost amplitude of the studied mode , and $ \hat{D} $ the dipole operator in our case, defined as
\begin{eqnarray}{\label{dipole}}
	\hat{D} & = & \frac{NZ}{A} \bigg( \frac{1}{Z}\sum_{p=1}^{Z}\vec{r}_p - \frac{1}{N}\sum_{n=1}^{N}\vec{r}_n \bigg) \nonumber \\
	& = & \frac{NZ}{A}\bigg( \vec{R}_p - \vec{R}_n \bigg),
\end{eqnarray}
where $ \vec{r}_p $ and $ \vec{r}_n $ are the position vectors of each proton and neutron respectively, and where $ \vec{R}_p $ and $ \vec{R}_n $ are the position vectors of the centers of mass of the protons and neutrons respectively. In order to obtain the three dipole modes (x, y, z) at the same time, we apply a diagonal excitation. For one mode (x-mode for example), the excitation is proportional to $e^{x} $. During the propagation of the TDHF equations in the time \cite{flocard1978}, the dipole moment $ D(t) = \sum_{i} \bra{\psi_{i}(t)}\hat{D}\ket{\psi_{i}(t)} $ is recorded in the time domain. Finally, the Fourier transform of the signal D (t) in the frequency domain gives the spectral distribution of the dipole strength $ S_{D}(\omega)$ given by \cite{ring1980}
\begin{eqnarray}{\label{strength}}
	S_{D}(\omega) & = & \sum_{\nu} \delta(E-E_{\nu}) \big \vert \bra{\nu}\hat{D}\ket{0}\big\vert^2. 
\end{eqnarray}
In order to the signal vanishes at the end of the simulation time,
some filtering is necessary to avoid artifacts in the spectra by cutting the signal at a certain time \cite{william1992}. In practice we use windowing in the time domain by damping the signal $ D(t) $  at the final time with $ cos \big(\frac{\pi t}{2T_{f}}\big)^n $ \cite{sky3d}.
\begin{equation}{\label{wind_d(t)}}
	D(t) \longrightarrow D_{fil} = D(t). cos \bigg(\frac{\pi t}{2T_{f}}\bigg)^n,
\end{equation}
where n represents the strength of filtering and $ T_{f} $ is the final time  of the simulation. In this work, we chose n = 6 which is sufficient to suppress the artifacts safely. More details can be found in Refs. \cite{sky3d, reinhard2006} . We kept the same number of grid points (24) and grid spacing (1 fm) as in static calculation. We chose nt= 4000 as a number of time steps to be run, and dt = 0.2 fm/c is the time step, so T$ _ {f}$ = 800 fm/c.
\section{Results and Discussion}\label{sec4}
\subsection{Ground-state properties of $^{92-108}\text{Mo}$ isotopes}
\begin{figure}[!ht]
	\begin{center}
		\includegraphics[width=0.8\textwidth]{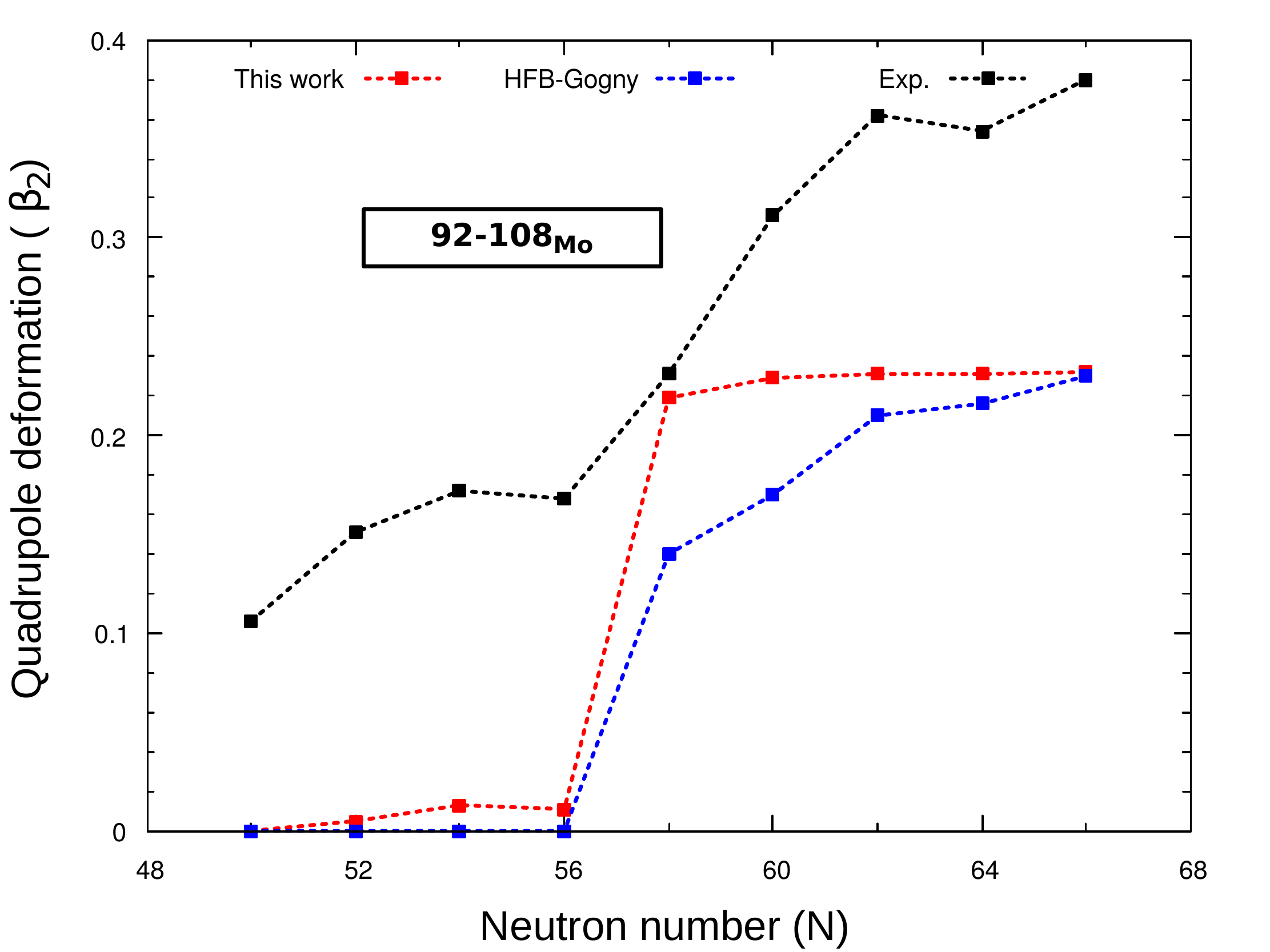}
		\caption{(Color online) The quadrupole deformation $ \beta_{2} $ of $ ^{92-108}\text{Mo} $ nuclei as function of the neutron number N compared with experimental data from Ref.\cite{raman2001} and theoretical data from \cite{delaroche2010}.}
		\label{b2-n}
	\end{center}
\end{figure}

\qquad The isotopic chain of the double-even $^{92-108}\text{Mo}$ under investigation  lies in the region that knows a transition between spherical, where the neutrons number N is close to the magic number 50, and deformed nuclei  when N increases \cite{carlos1974,yoshida2011,maruhn2005}. The deformation in this region is mainly due to the filling of the N = 50 shell gap. The static calculations, described previously, were performed with SLy6 parametrization \cite{CHABANAT1998}. BCS pairing is included using the volume-$\delta$ interaction (VDI) \cite{sky3d}, with pairing strengths $ V_{p} $ = 298.760 MeV and $ V_{n} $ = 288.523 MeV for protons and neutrons respectively.  The two  deformation parameters $ \beta_{2} $, and $ \gamma $ 
are among the most important ground-state properties, which give us an idea about the shape of the nucleus \cite{ring1980,takigawa2017}. They are treated as a probe to select ground-states of all nuclei under study.
In Table \ref{tab2}, we summarize the numerical results for the couple ($\beta_{2}$,$\gamma$)  of $ ^{92-108}\text{Mo} $ isotopes, including the available experimental data from Ref.\cite{raman2001} and the HFB calculations based on the D1S Gogny force~\cite{delaroche2010} for comparison. In Fig.\ref{b2-n}, we plotted the variation of $ \beta_{2} $ as a function of neutrons number (N). It should be noted that Fig.\ref{b2-n} does not indicate the sign of the quadrupole deformation $\beta_{2}$.
\begin{table}[!ht]
	\centering
	\caption{The  deformation parameters ($\beta_{2}$,$\gamma$) for Mo isotopes calculated with SLy6 parametrization are compared with the experimental data extracted from Ref.\cite{raman2001}, and data from Ref.\cite{delaroche2010}. \label{tab1} }
	{\begin{tabular}{@{}ccccccc@{}} \hline \hline
			Nucleus &  SLy6 &&  HFB$\_$Gogny.\cite{delaroche2010}&&Exp.\cite{raman2001}\\
			\hline
			$ ^{92}\text{Mo} $  & (0.000; ---) && (0.000; $0.0^\circ$) &&  (0.106; ---)  \\
			$ ^{94}\text{Mo} $  &  (0.005; ---)&& (0.000; $0.0^\circ$) &&  (0.151; ---)  \\
			$ ^{96}\text{Mo} $  & (0.013; $0.0^\circ$) && (0.000; $0.0^\circ$) &&  (0.172; ---)  \\
			$ ^{98}\text{Mo} $  & (0.011; $0.0^\circ$) && (0.000; $0.0^\circ$) &&  (0.168; ---)  \\
			$ ^{100}\text{Mo} $ & (0.219; $60^\circ$)  && (0.140; $44^\circ$)  &&  (0.231; ---)  \\
			$ ^{102}\text{Mo} $ & (0.229; $60^\circ$)  && (0.170; $58^\circ$)  &&  (0.311; ---)  \\
			$ ^{104}\text{Mo} $ & (0.231; $57^\circ$)  && (0.210; $60^\circ$)  && (0.362; ---)  \\
			$ ^{106}\text{Mo} $ & (0.231; $60^\circ$)  && (0.216; $59^\circ$)  &&  (0.354; ---)  \\
			$ ^{108}\text{Mo} $ & (0.232; $60^\circ$)  && (0.230; $60^\circ$)  &&  (0.380; ---)  \\
			\hline \hline
	\end{tabular}}
\end{table}

From Fig.\ref{b2-n}, we can clearly see an agreement between our calculations within the HF plus BCS theory based on Skyrme force (SLy6 in our case) and those of the HFB theory based on D1S Gogny force \cite{delaroche2010}. In recent publication, G.Colo et al.\cite{colo2020} have  published a work on ISGMR and ISGQR in $ ^{92-100}\text{Mo} $ using QRPA approach based on Skyrme interactions. This study includes static constrained calculations with different Skyrme forces (SLy6 and others) shown in Fig.1. We limit our comparison to Figure 1(c) for SLy6 that we used in this work.
They have found, in overall, an increasing of softness as well as the spherical minimum becomes more and more shallow as the number of neutrons N increases.
In $ ^{92}\text{Mo}$, a spherical minimum is observed which confirms  that it has an approximately spherical shape as we have predicted  ($\beta_2 = 0$). In $ ^{94,96}\text{Mo}$, the spherical minimum becomes  shallow, especially in $ ^{96}\text{Mo}$, which is roughly in agreement with our results ($ ^{94}\text{Mo}$ is nearly spherical and $ ^{96}\text{Mo}$ is weakly deformed). In the neutron-rich $ ^{94,96}\text{Mo}$ isotopes, the potential energy curve (PEC) $ ^{94,96}\text{Mo}$ is very shallow. In particular, $ ^{100}\text{Mo}$ presents  a kind of convex curve with a rather modest barrier at $\beta_2 \simeq0$. In our calculations, we found two local minima in $ ^{100}\text{Mo}$ (one oblate and the other triaxial) as we will see later.\\
On the other hand, the experimental results \cite{raman2001} are a bit different from our calculations, but we reproduce the overall trend of increasing the quadrupole distortion $ \beta_{2} $ as the  neutrons number N increases from the magic number N = 50. We point out that one should be careful in direct comparison between theoretical $\beta_2$ and "experimental" one, especially for spherical nuclei, which is extracted from actual experimental data ( usually B(E2) transition strength) under certain model assumptions including that the nucleus is not spherical.

According to the values of the couple ($\beta_{2}$,$\gamma$), we can predict the shape of nuclei. From Table \ref{tab1}, the semi-magic nucleus $ ^{92}\text{Mo}$ (N=50) has a spherical shape where $\beta_{2} \simeq 0$. For the three nuclei $ ^{94}\text{Mo}$, $ ^{96}\text{Mo}$ and $ ^{98}\text{Mo}$, the value of $\beta_{2}$ deviates slightly from 0, so they have an "approximate" spherical form. As we reach N = 58 ($ ^{100}\text{Mo}$), there is transition from spherical to permanently deformed ground state with oblate shape ($\gamma = 60 ^\circ$). We point out that we have not indicated the triaxiality parameter $\gamma$ for spherical nuclei $ ^{92-94}\text{Mo} $ because it is not defined for $\beta_2 =0$ and therefore his inclusion is meaningless.\\ 

The neutron-rich nuclei with A$\sim$100 are known by instability of shapes, which can lead to a shape coexistence \cite{stachel1982,wrzosek2012}. For the nucleus $ ^{100}\text{Mo}$, we found two minima in energy of the ground state as indicated in Table \ref{tab2}. The difference in energy $\Delta E$ between oblate and triaxial minima is very small of the order of 0.05 MeV. This is a clear indication of a shape coexistence for this nucleus. We will treat the two minima in dynamic calculation in the next section.
\begin{table}[!ht]
	\centering
	\caption {The ground-state properties of two minima for $^{100}\text{Mo}$ nucleus.  \label{tab2}} 
	{\begin{tabular}{@{}ccccccccc@{}} \hline \hline
			Properties&& Oblate minimum &&& Triaxial minimum&&&  \\
			\hline
			Binding energy (B.E) && -857.52 MeV&&&  -857.47 MeV&&& \\
			Root mean square (r.m.s)  && 4.467 fm&&&  4.470 fm&&&  \\
			Quadrupole deformation $\beta_{2}$  && 0.219 &&& 0.242&&&  \\
			Triaxiality parameter $\gamma$ && $60^\circ$ &&& $23^\circ$&&&  \\
			\hline \hline
	\end{tabular}}
\end{table}
\subsection{GDR in $ ^{92-108}\text{Mo} $ isotopes }

\begin{figure}[!ht]
	\centering
	\includegraphics[width=1.0\textwidth]{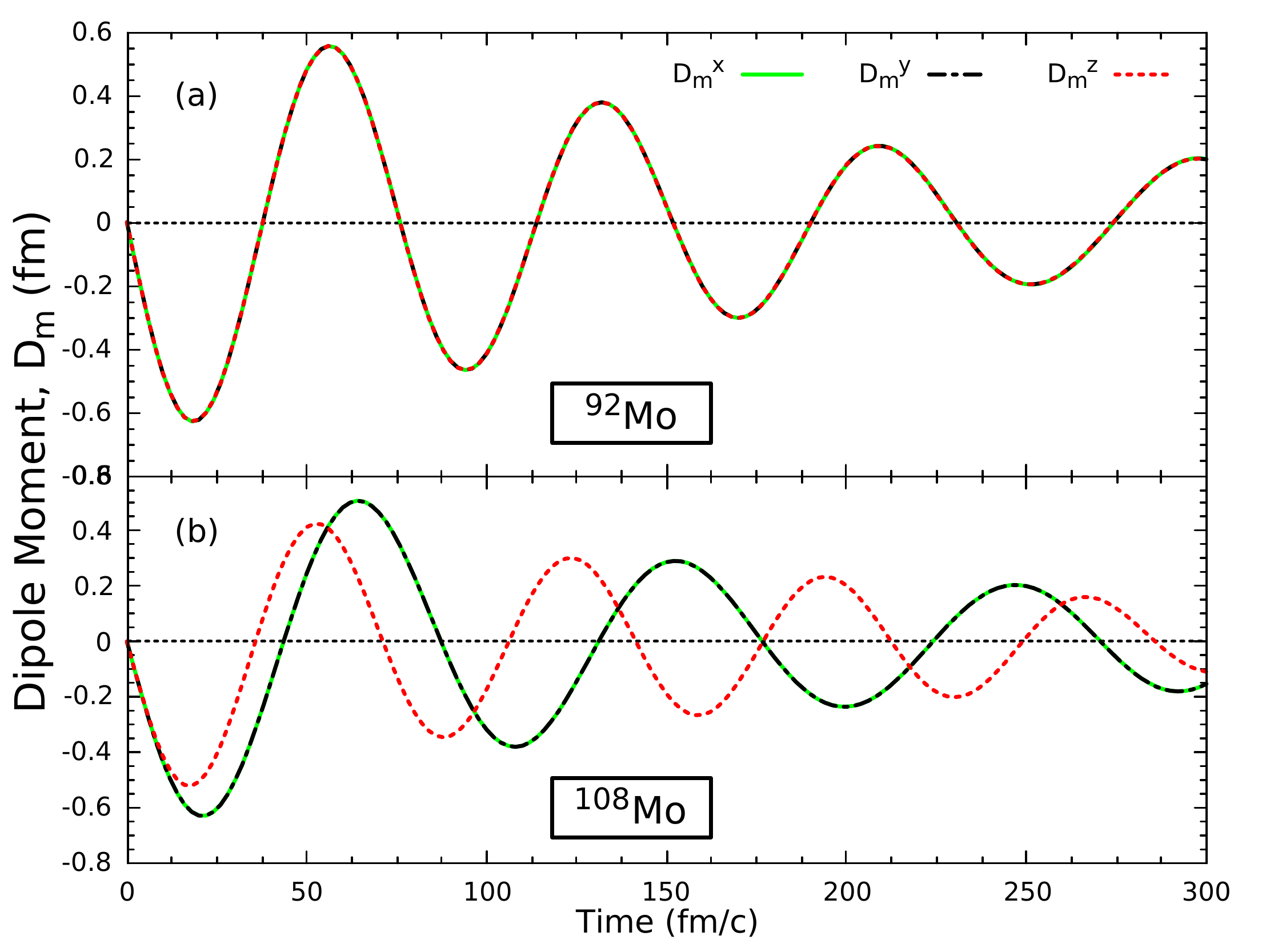} 
	\caption{(Color online) The three components of dipole moment $ D_{m}^{i}(t) $ plotted as function of the simulation time  t(fm/c), are calculated with SLy6 parametrization for: (a) $^{92}\text{Mo}$, and (b) $^{108}\text{Mo}$ nuclei.}
	\label{dm-t1}
\end{figure}

\begin{figure}[!ht]
	\centering
	\includegraphics[width=1.0\textwidth]{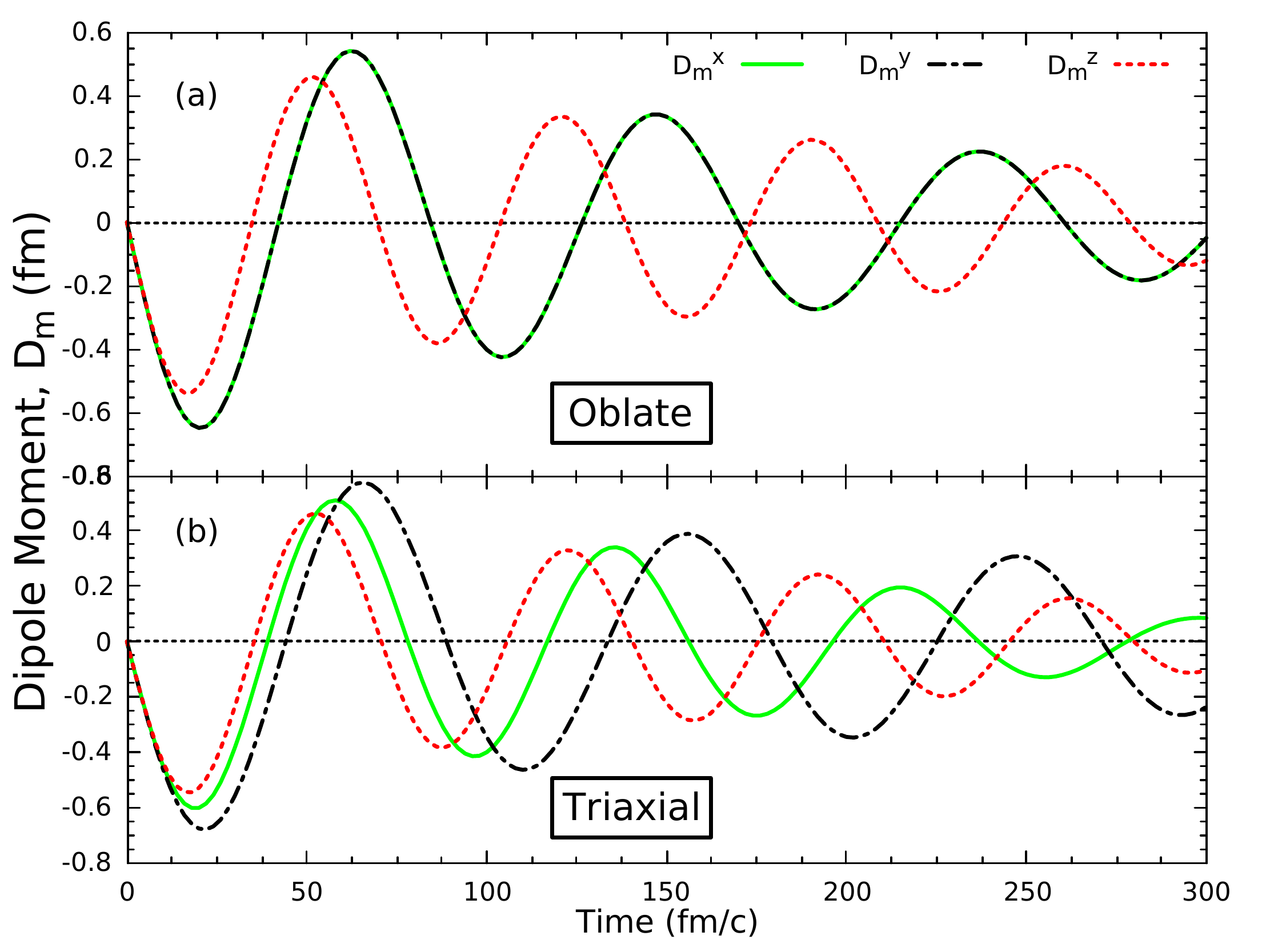} 
	\caption{(Color online) The same as Fig.\ref{dm-t1} for $^{100}\text{Mo}$ nucleus: (a) oblate, and (b) triaxial shape.}
	\label{dm-t2}
\end{figure}

\qquad The response of the nucleus under study to the isovector dipole boost (Eq.\ref{boost}) is expressed by the evolution of the dipole moment $D_{m}(t)$ (Eq.\ref{dipole}) in time domain. The three components $D^{i}_{m}(t)$, where i = x, y, z, can inform us on the collective motions of nucleons along the three directions x, y and z.
Figures \ref{dm-t1}-\ref{dm-t2} display the time evolution of $D^{i}_{m}(t)$ for the nuclei $^{92}\text{Mo}$, $^{100}\text{Mo}$ and $^{108}\text{Mo}$. We chose only three nuclei among nine ones (one spherical, one oblate, and one triaxial).

In Fig.\ref{dm-t1}a, due to spherical shape for $^{92}\text{Mo}$ as predicted in static calculation, the three components of dipole moment ($D_{m}^{x}, D_{m}^{y}, D_{m}^{z}$) are identical, i.e., the oscillation frequencies along the three axes are equal $\omega_{x}=\omega_{y}=\omega_{z}$. The periodicity of $D^{i}_{m}(t)$ allows us to estimate the excitation energies  E$_{i}$ of the oscillations along each of the three axes x, y and z. For each $D^{i}_{m}(t)$, the period of T $\approx 75.5$ fm/c gives an excitation energy of about E$_{i}\approx$ 16.42 MeV, which is very close to the experimental value $ E_{GDR}^{exp.}$ = 16.9$\pm$0.1 MeV \cite{beil1974}. In Fig.\ref{dm-t1}b, the $D^{z}_{m}(t)$ values differ from $D^{x}_{m}(t)$ and $D^{y}_{m}(t)$ values which are identical, i.e., $\omega_{x}=\omega_{y}\ne \omega_{z}$. This was expected for an axially deformed nucleus $^{108}\text{Mo}$. For $D^{z}_{m}(t)$ (resp. $D^{x}_{m}(t)$), the period is of the order of 71.64 fm/c (resp. 87.62 fm/c), which corresponds to resonance energy of E$_{z}\approx$ 17.30 MeV (resp. E$_{x}\approx$ 14.15 MeV). Since E$_{z}$ $ > $ E$_{x}$ = E$_{y}$ and due to relation (\ref{E_R-1}), the radius $R_{z} <  R_{x} = R_{y}$ shows that $^{108}\text{Mo}$ has an oblate shape as expected in static calculation ($\gamma =60^\circ$ )


\begin{figure}[!htp]
	\centering
	\includegraphics[width=1.0\textwidth]{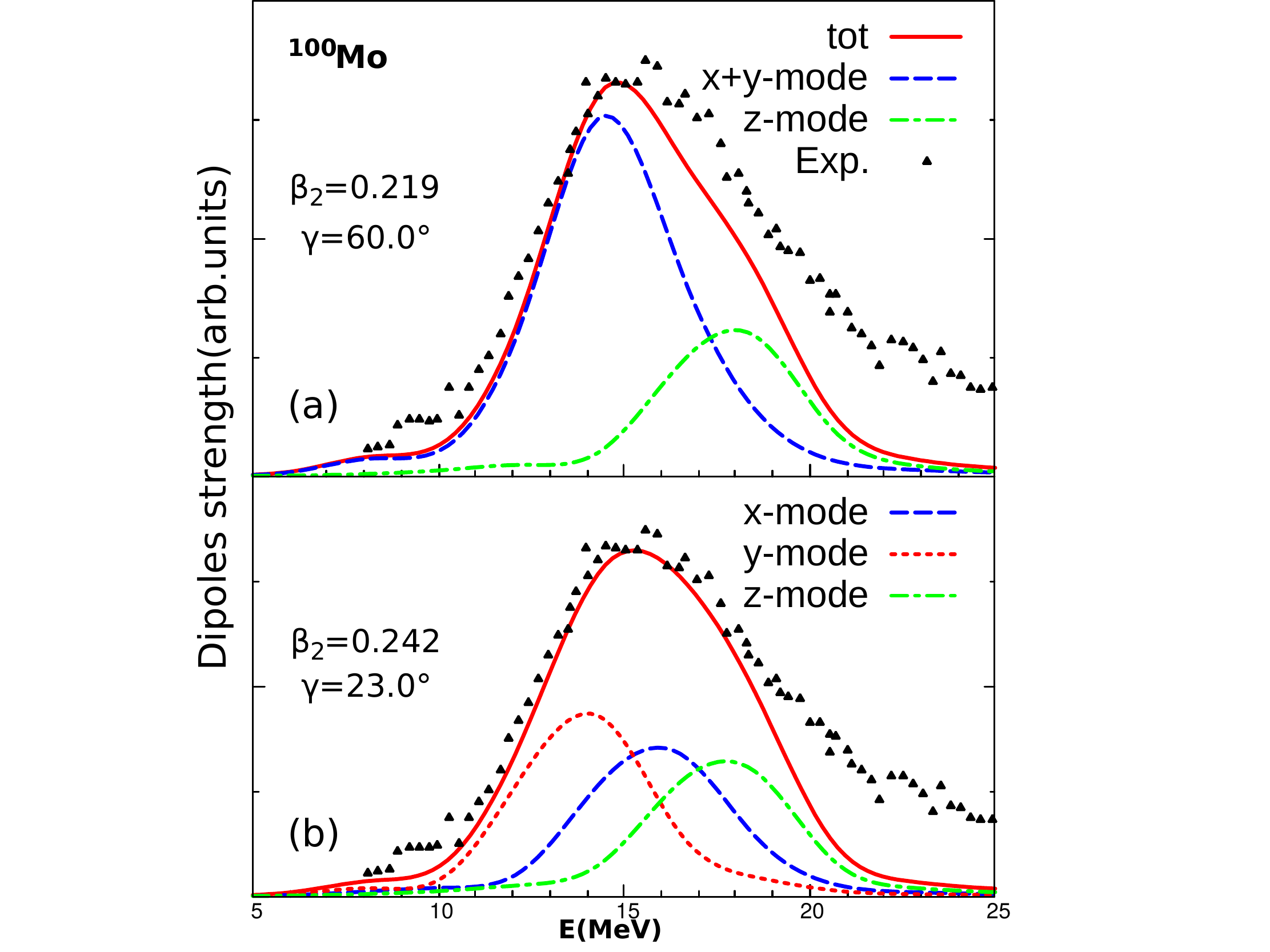} 
	\caption{(Color online) The calculated GDR spectra for $^{100}\text{Mo}$ with the Skyrme force SLy6: oblate (a) and triaxial (b). The experimental data are extracted from \cite{beil1974}}
	\label{ob_tr}
\end{figure}

Fig.\ref{dm-t2} shows the time evolution  of D(t) for the    $ ^{100}\text{Mo} $ nucleus. Fig.\ref{dm-t2}a presents the same case as in Fig.\ref{dm-t1}a, where $ ^{100}\text{Mo} $ has an oblate shape with  resonance energies along the axes z and x been respectively E$_{z}$ $\approx$ 17.70 MeV and E$_{x}\approx$ 14.70 MeV .
In Fig.\ref{dm-t2}b, we notice that the oscillation frequencies $\omega_{i}$ along the three axes  are different from each other $\omega_{x}\neq \omega_{y}\neq \omega_{z}$,which is expected for triaxial nucleus. The estimated resonance energies E$_{i}$ along the three axes x, y, z are respectively E$_{x}$ $\approx$ 15.84 MeV, E$_{y}$ $\approx$ 13.90 MeV, E$_{z}$ $\approx$ 17.58 MeV.\\
Fig.\ref{ob_tr} presents the calculated GDR spectra corresponding to two minima (triaxial, oblate) together with the available experimental data. It confirms this suggestion: the upper panel (Fig.\ref{ob_tr}a) shows an oblate shape for $^{100}\text{Mo}$ due to oscillations along the shorter axis z (K=0 mode) which are characterized by higher energies than the oscillations along the longer axis x and y ($ |K|=1 $ mode) perpendicular to it, i.e, $E_{x}= E_{y} < E_{z}$, while the lower panel (Fig.\ref{ob_tr}b) shows a triaxial shape characterized by three different GDR peaks that correspond to the three different principal axes x, y and z with resonance energies $E_{x}\neq E_{y}\neq E_{z}$. On the other hand, we can see an overall agreement between the both total strengths (oblate and triaxial) and experimental data, with a slight advantage in the case of triaxial shape. Thus, one can predict a static triaxial deformation in the ground state of $^{100}\text{Mo}$. In recent publication \cite{shi2022}, the authors have found, in the light of studying ISGMR in $^{100}\text{Mo}$, that it is triaxially deformed in the ground state.

\begin{figure}[!ht]
	\begin{center}
		\begin{minipage}[t]{0.46\textwidth}
			\includegraphics[scale=0.7]{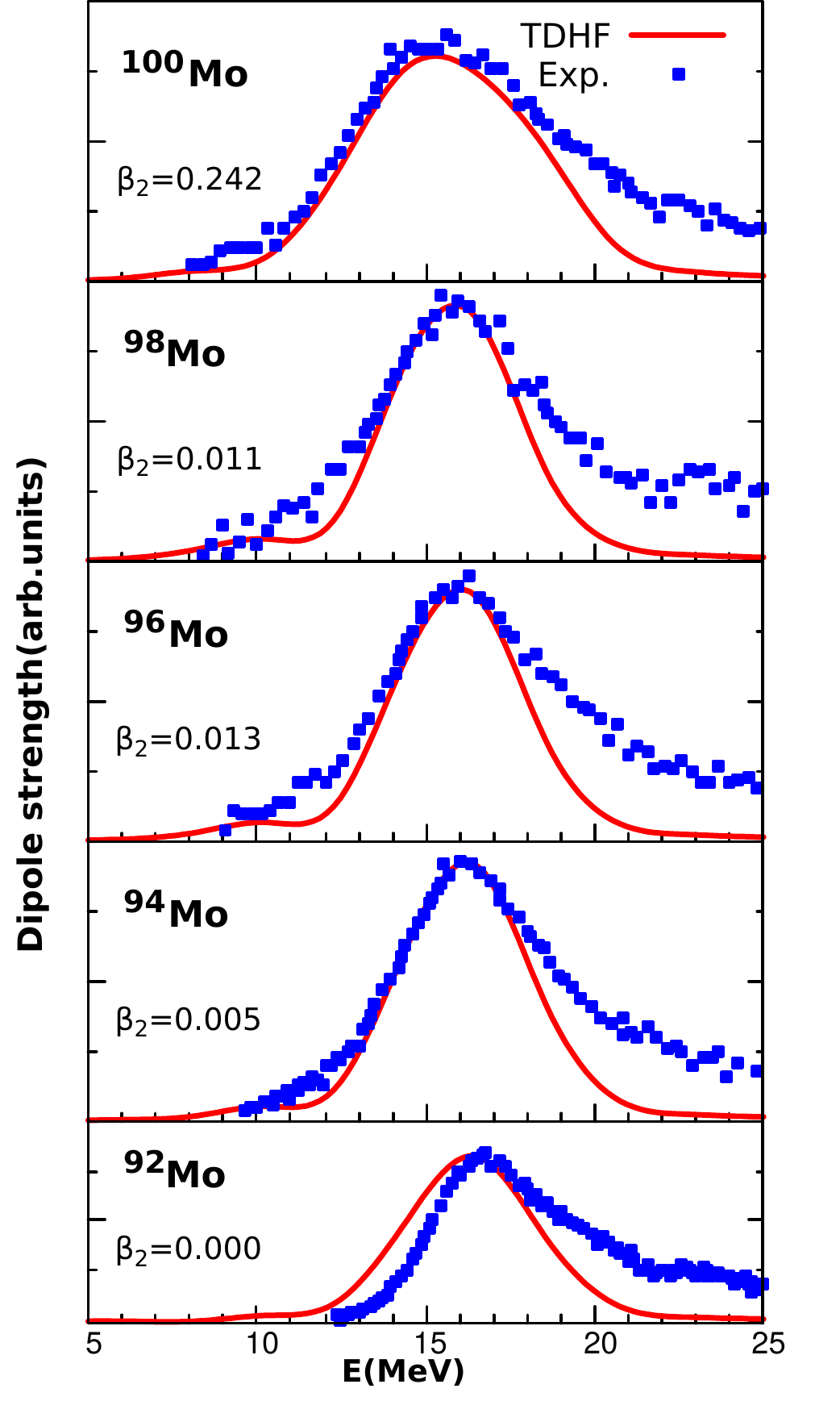}
		\end{minipage}
		\vspace{0.5cm}
		\hspace{0.5cm}
		\begin{minipage}[t]{0.46\textwidth}
			\includegraphics[scale=0.7]{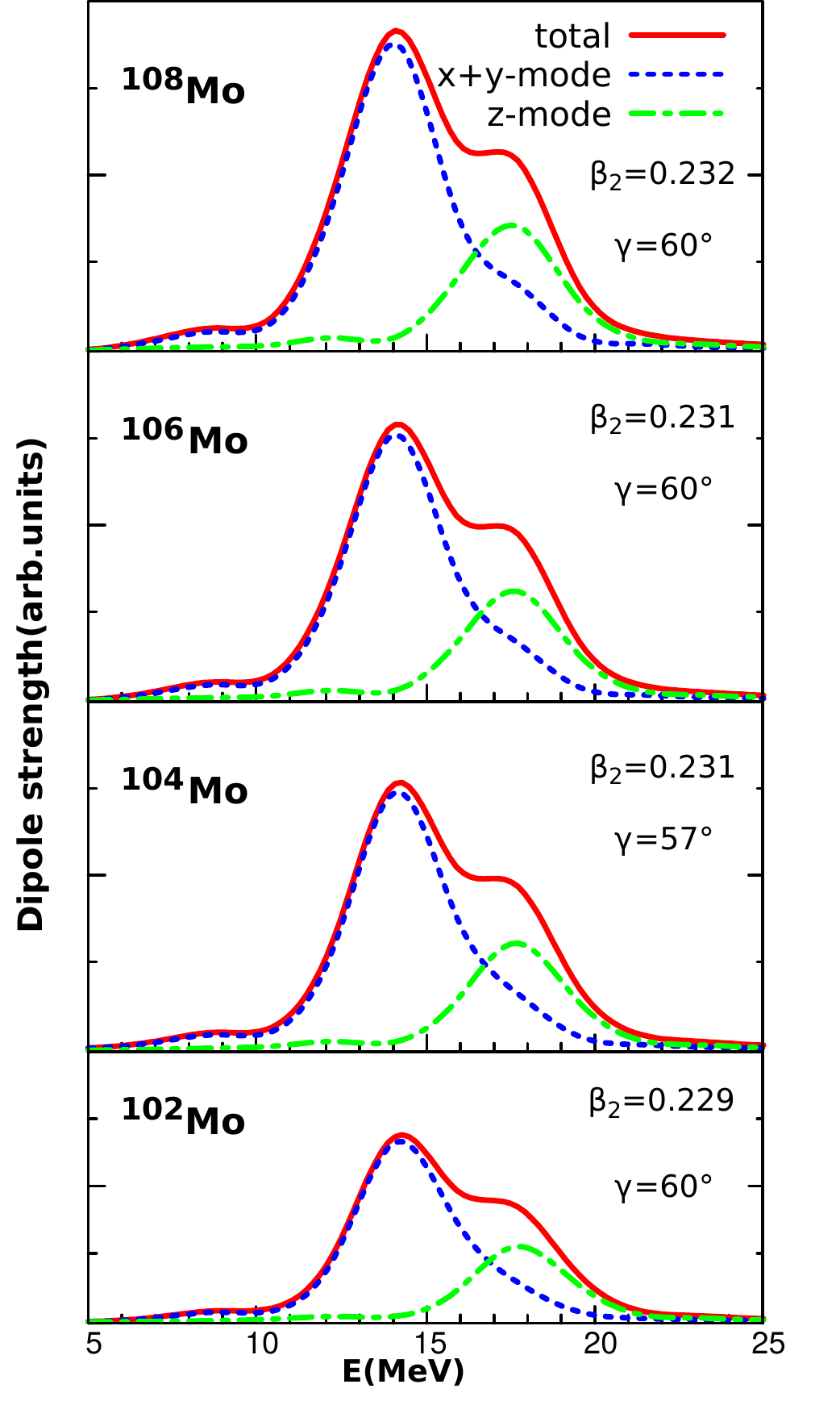}
		\end{minipage}
		
		\caption{ (Color online) GDR strengths in the chain of $ ^{92-108}\text{Mo} $ calculated with SLy6 parametrization. Left side: our calculations (red line) are compared with experimental data (blue square) from Ref.\cite{beil1974}. Right side:	The solid(red), dashed(blue) and dotted-dashed(green) lines denote the dipole strengths: total, along the long axis and the short axis(multiplied by 2) respectively.} 
		\label{gdr92-100}
	\end{center}	
\end{figure}

In order to obtain the GDR strengths S(E) (\ref{strength}), we calculated the Fourier transform of the isovector signal D(t) which is  simply the imaginary part of the Fourier transform of D(t)\cite{maruhn2006}.\\
Fig.\ref{gdr92-100} displays the GDR strengths  in $^{92-108}\text{Mo}$ isotopes computed with the Skyrme forces SLy6, and compared with the available experimental data from Ref.\cite{beil1974}. In the left part of Fig.\ref{gdr92-100}, we can generally see an agreement between our calculations (TDHF) and the experimental data. The agreement is better for the deformed nucleus $^{100}\text{Mo}$. We notice also that the GDR width $\Gamma$ calculated with TDHF is always underestimated because the contribution $\Gamma^{\downarrow}$ of the total width $\Gamma$ caused by 2p-2h (and higher) collisions is not covered under TDHF theory, this requires to go beyond mean-field \cite{maruhn2005}. The right side of Fig.\ref{gdr92-100} shows the GDR strengths predicted in $^{102-108}\text{Mo}$ nuclei. We plotted also the k = 0 and k = 1 modes, along z-axis and x+y axes respectively, to illustrate the magnitude and oblate character of the deformation. It is seen that  $^{102-108}\text{Mo}$ nuclei have GDR spectra corresponding to an oblate shape, where the component (K = 0) along z-axis has a high resonance energy compared to that (K = 1) along the x- and y-axes, i.e, $E_{z} > E_{x} = E_{y}$, but the intensity of the lower component is about twice that of the upper component due to the fact that the vibrations along equivalent axes (x and y in this case) are degenerated \cite{Masur2006}. On the other hand, as the neutron number increases from spherical $^{92}\text{Mo}$ to $^{108}\text{Mo}$ deformed nucleus, we notice that the width $\Gamma$ of GDR is related to deformation parameter $\beta_{2}$ as indicated on the right of two panels in Fig.\ref{gdr92-100}.

\begin{figure}[!ht]
	\centering
	\includegraphics[width=1.0\textwidth]{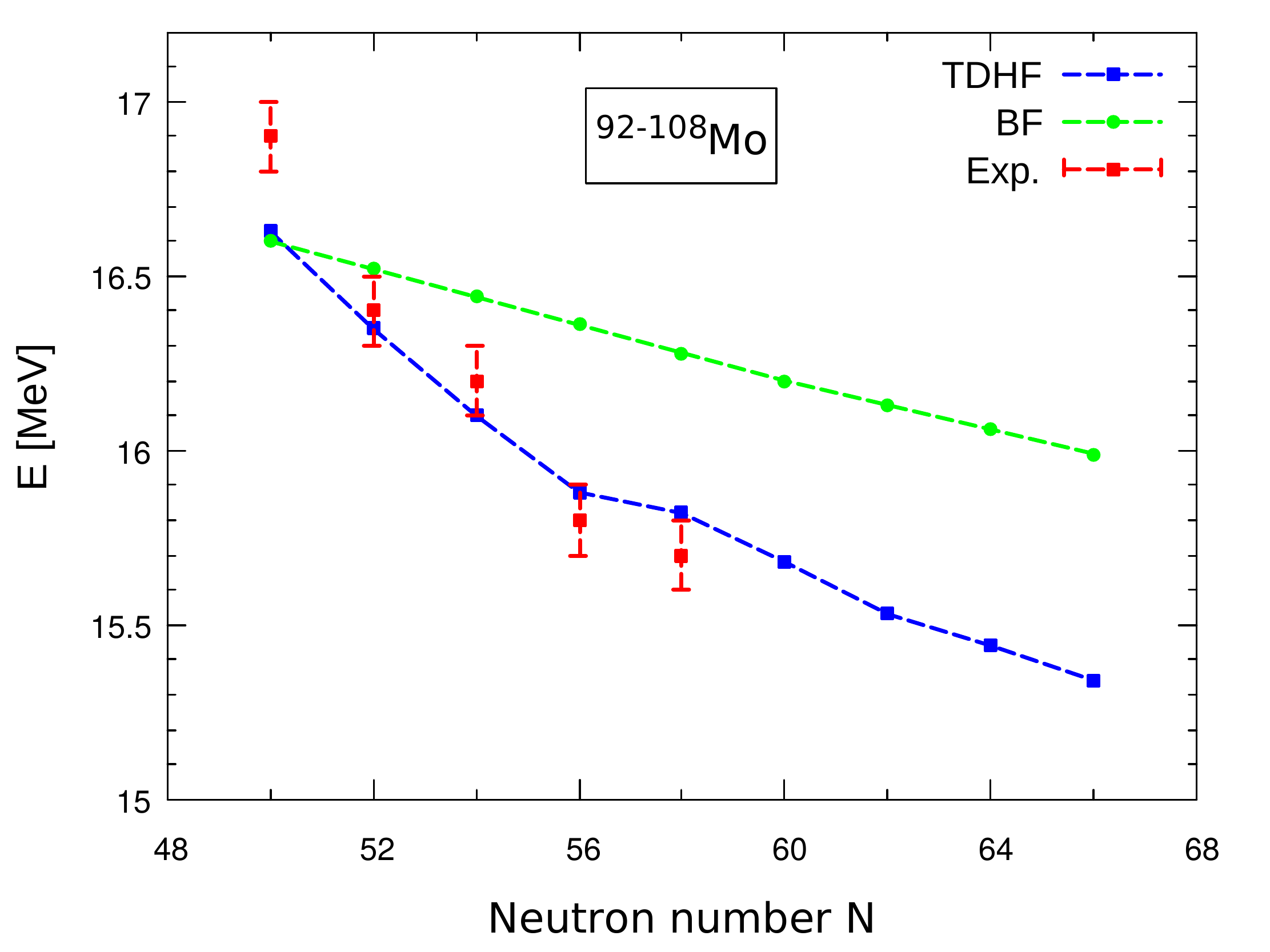} 
	\caption{(Color online) The IVGDR energies as a function of the neutron number N for $^{92-108}\text{Mo}$ nuclei. Comparison between our calculations (TDHF), experimental data \cite{beil1974} and the semi-empirical BF formula [Eq.\ref{BF}].}
	\label{E&N}
\end{figure}

From the calculated strengths GDR, we can estimate some resonance characteristics like  centroid energy $ E $, deformation splitting $\Delta E$ and width $\Gamma$, and the trends of these characteristics with the mass number A. In deformed nuclei, GDR strength splits into two components (k=0, k=1), which each of them has resonance energy $ E_{i}, i=1,2$. An estimate of the  energy centroid $ E $ is given by \cite{garg2018}
\begin{equation}{\label{centroid_energy}}
	E = \frac{\int_{0}^{+\infty} S(E)E dE}{\int_{0}^{+\infty} S(E) dE},
\end{equation} 
where S(E) (\ref{strength}) represents the GDR strength function.\\
Table \ref{resonance_energy} shows the calculated resonance energy $ E $ with SLy6 parametrization, compared with the available experimental data from Ref.\cite{beil1974}, and the empirical estimates based on Berman-Fultz (BF) model defined as \cite{berman1975,van1987}
\begin{equation}{\label{BF}}
	E = (31.2 A^{-1/3} + 20.6 A^{-1/6}) MeV
\end{equation} 
The Berman-Fultz model takes into account both surface  and volume contributions and treats GDR as a combination of Goldhaber-Teller\cite{goldhaber1948} ($ E \sim A^{-1/6}$) and Steinwedel-Jensen\cite{steinwedel1950} ($ E \sim A^{-1/3}$) models. In Fig.\ref{E&N}, we plotted the resonance energy as a function of the neutron number N. It can be seen that our results generally agree well with experimental data, and we reproduce the overall trend of decreasing resonance energy with increasing mass number A as shown by the sloped line obtained from the BF formula  (Eq.\ref{BF}).

\begin{table}[!ht]
	\centering
	\caption{The  calculated resonance energies ($E_{sly6}$) for $ ^{92-108}\text{Mo}$ nuclei are compared with the experimental data ($E_{exp.} $) extracted from Ref.\cite{beil1974}, and the estimates ($E_{BF}$) [Eq.\ref{BF}]. \label{resonance_energy} }
	{\begin{tabular}{@{}ccccccc@{}} \hline \hline
			Nucleus &           $E_{sly6}(MeV)$ &&  $E_{BF}$(MeV)[Eq.\ref{BF}]&&$ E_{exp.}$(MeV)\cite{beil1974}\\
			\hline
			$ ^{92}\text{Mo} $  & 16.63 && 16.60 &&  16.90 $\pm$ 0.1  \\
			$ ^{94}\text{Mo} $  & 16.35&& 16.52 &&  16.40 $\pm$ 0.1  \\
			$ ^{96}\text{Mo} $  & 16.10 && 16.44 &&  16.20 $\pm$ 0.1  \\
			$ ^{98}\text{Mo} $  & 15.88 && 16.36 &&  15.80 $\pm$ 0.1  \\
			$ ^{100}\text{Mo} $ & 15.82  && 16.28  &&  15.70 $\pm$ 0.1  \\
			$ ^{102}\text{Mo} $ & 15.68  && 16.20  &&  ---  \\
			$ ^{104}\text{Mo} $ & 15.89  && 16.13  && ---  \\
			$ ^{106}\text{Mo} $ & 15.44  && 16.06  &&  ---  \\
			$ ^{108}\text{Mo} $ & 15.34  && 15.99  &&  ---  \\
			\hline \hline
	\end{tabular}}
\end{table}

\begin{figure}[!ht]
	\centering
	\includegraphics[width=1.0\textwidth]{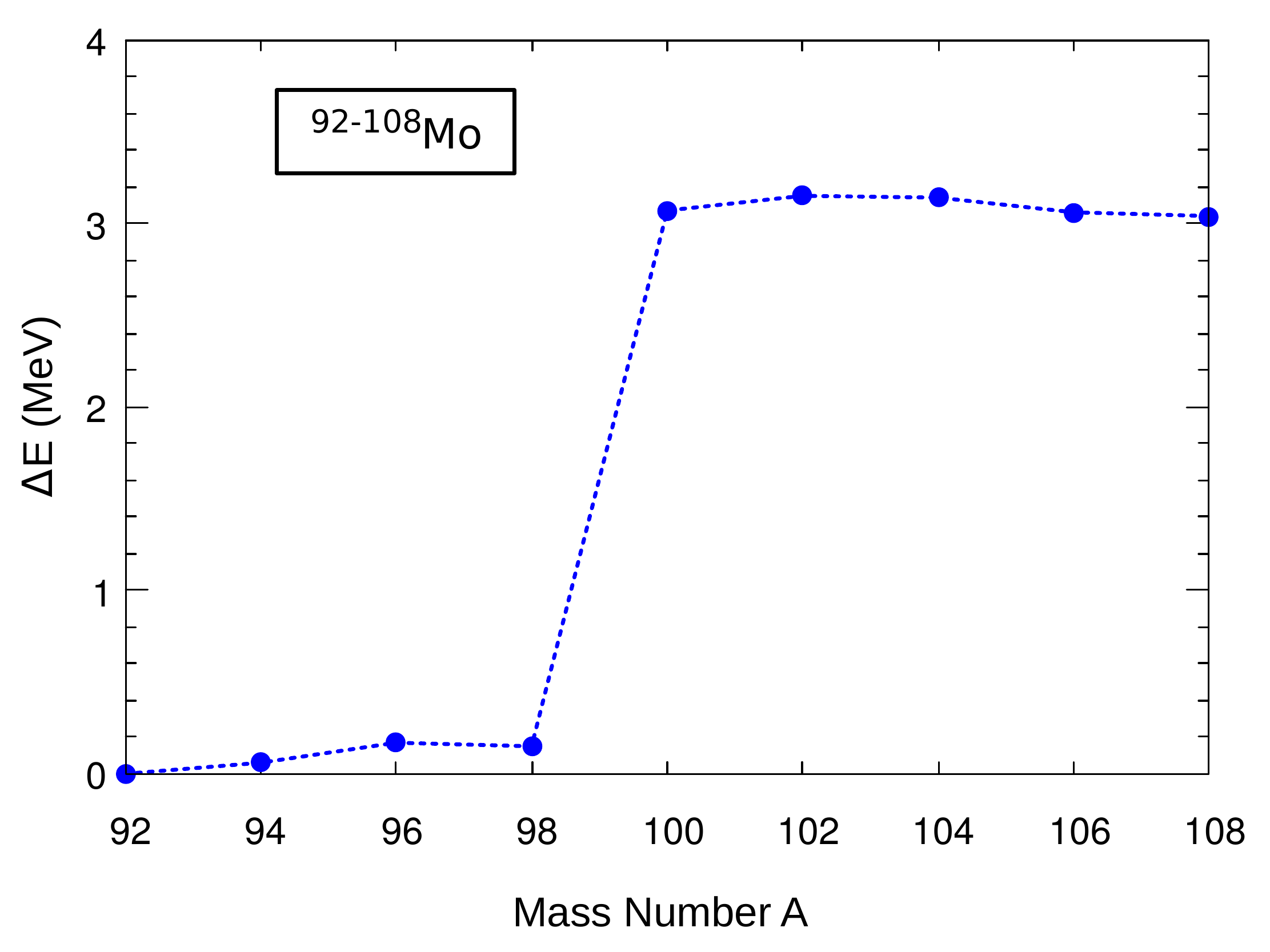} 
	\caption{(Color online) The deformation splitting $\Delta E$ of $^{92-108}\text{Mo}$ nuclei as a function of the mass number A  calculated with SLy6 parametrization.}
	\label{delta_E}
\end{figure}

The deformation splitting $\Delta E = E_{2} - E_{1}$ represents the "distance" in MeV between two  peak positions of GDR strength in deformed nuclei. It can give an idea about the magnitude of deformation. Fig.\ref{delta_E} displays the deformation splitting $\Delta E$ as a function of the mass number A of Mo isotopes. It should be noted that the resonance energies $ E_{1} $ and $E_{2}$ are calculated by using the relation (\ref{centroid_energy}). It is seen that $\Delta E$ is too small for  $^{94-98}\text{Mo}$ nuclei, which confirms that these nuclei are very weakly deformed. It vanishes ($\Delta E = 0$) for the semi-magic nucleus $^{92}\text{Mo}$ (N = 50), which is not  surprising because for a spherical nucleus the three peaks of GDR coincide, i.e, $ E_{x} = E_{y} = E_{z}$. From $^{98}\text{Mo}$ to $^{100}\text{Mo}$, there is a shape transition from an approximate spherical to permanently deformed nuclei, where the deformation splitting $\Delta E$ increases suddenly to $\Delta E \approx 3MeV$. It confirms that GDR-splitting is caused by the deformation structure of nuclei, i.e., $\Delta E \sim \beta_{2}$\cite{danos1958}.

In order to study the GDR dependence on nuclear matter properties of Skyrme forces such as asymmetry energy $a_{s}$,  we chose three Skyrme interactions  with $a_{s}$ different as shown in Table \ref{a_sym}.
\begin{table}[!ht]
	\centering
	\caption {The asymmetry energy $ a_{s} $ for different Skyrme forces.  \label{a_sym}} 
	{\begin{tabular}{@{}ccccccccc@{}} \hline \hline
			Forces &&  $ a_{s}(MeV) $  \\
			\hline
			SVbas \cite{reinhard2009} && 30.00  \\
			SLy6 \cite{CHABANAT1998} &&  32.00\\
			SkI3\cite{REINHARD1995}&& 34.80  \\
			\hline \hline
	\end{tabular}}
\end{table}
\begin{figure}[!ht]
	\centering
	\includegraphics[width=1.0\textwidth]{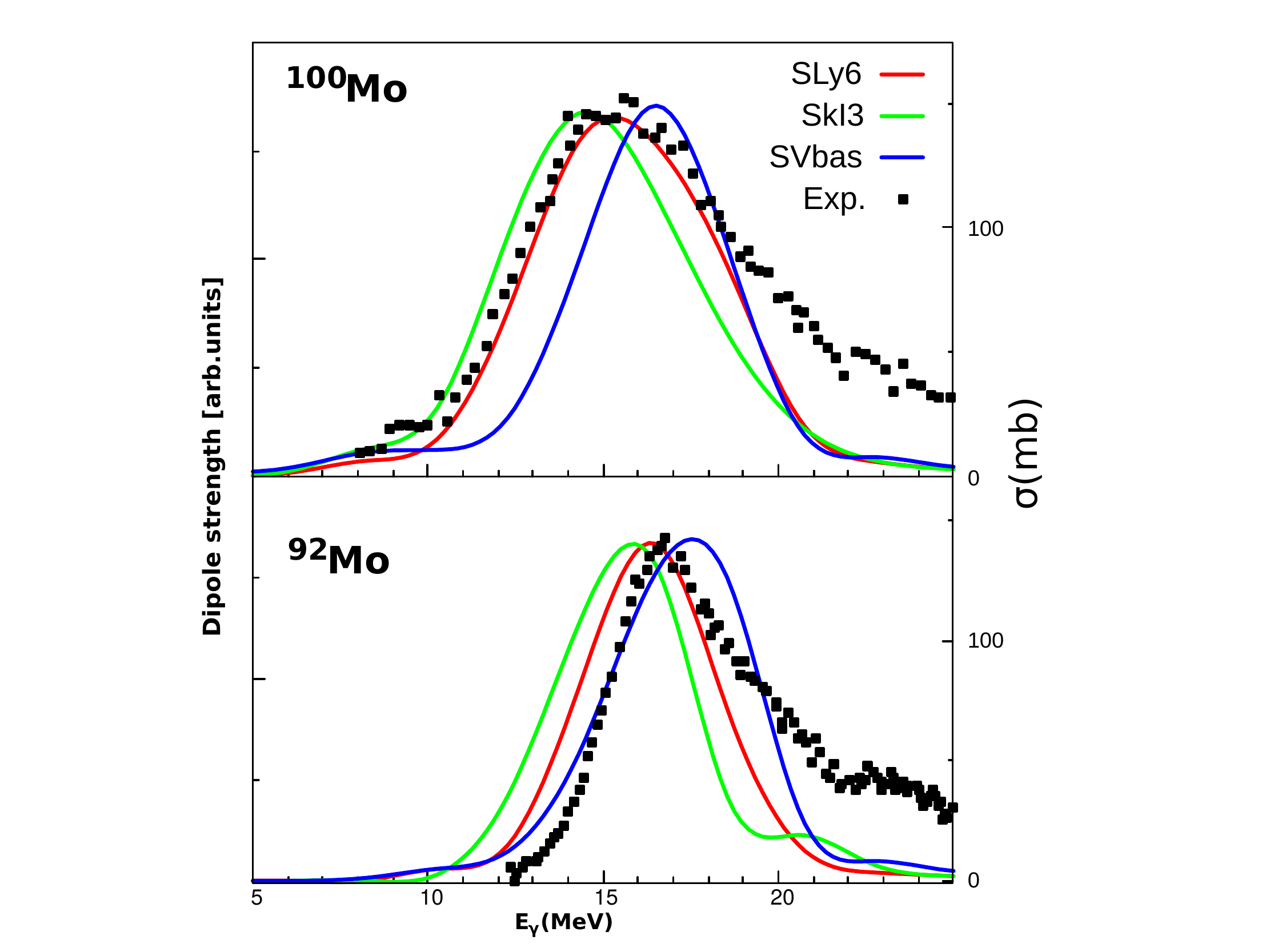} 
	\caption{(Color online) The GDR strengths for $^{92,100}\text{Mo}$ nuclei calculated with SLy6 (red line), SkI3 (green line), SVbas (blue line) parametrization, compared with the experimental data\cite{beil1974}.}
	\label{comp_gdr}
\end{figure}
Fig.\ref{comp_gdr} displays  GDR strength for $^{92}\text{Mo}$ and $^{100}\text{Mo}$ calculated with the Skyrme forces SLy6, SkI3 and SVbas, including experimental data \cite{beil1974} for comparison. We can see that there is a slight difference regarding the peak position  between these forces, with a shift of the GDR strength towards the higher energy for SVbas and the opposite occurs for Ski3. The Skyrme force SLy6 is the best among these forces which reproduces the experimental data. This can be clearly seen in Fig.\ref{comp_E} showing the resonance energy as a function of the neutron number N. One can see that all the forces reproduce the overall trend of decreasing energy with increasing N (so mass number A), with an advantage for SLy6 where their results are very closer to experimental data. 
\begin{figure}[!ht]
	\centering
	\includegraphics[width=1.0\textwidth]{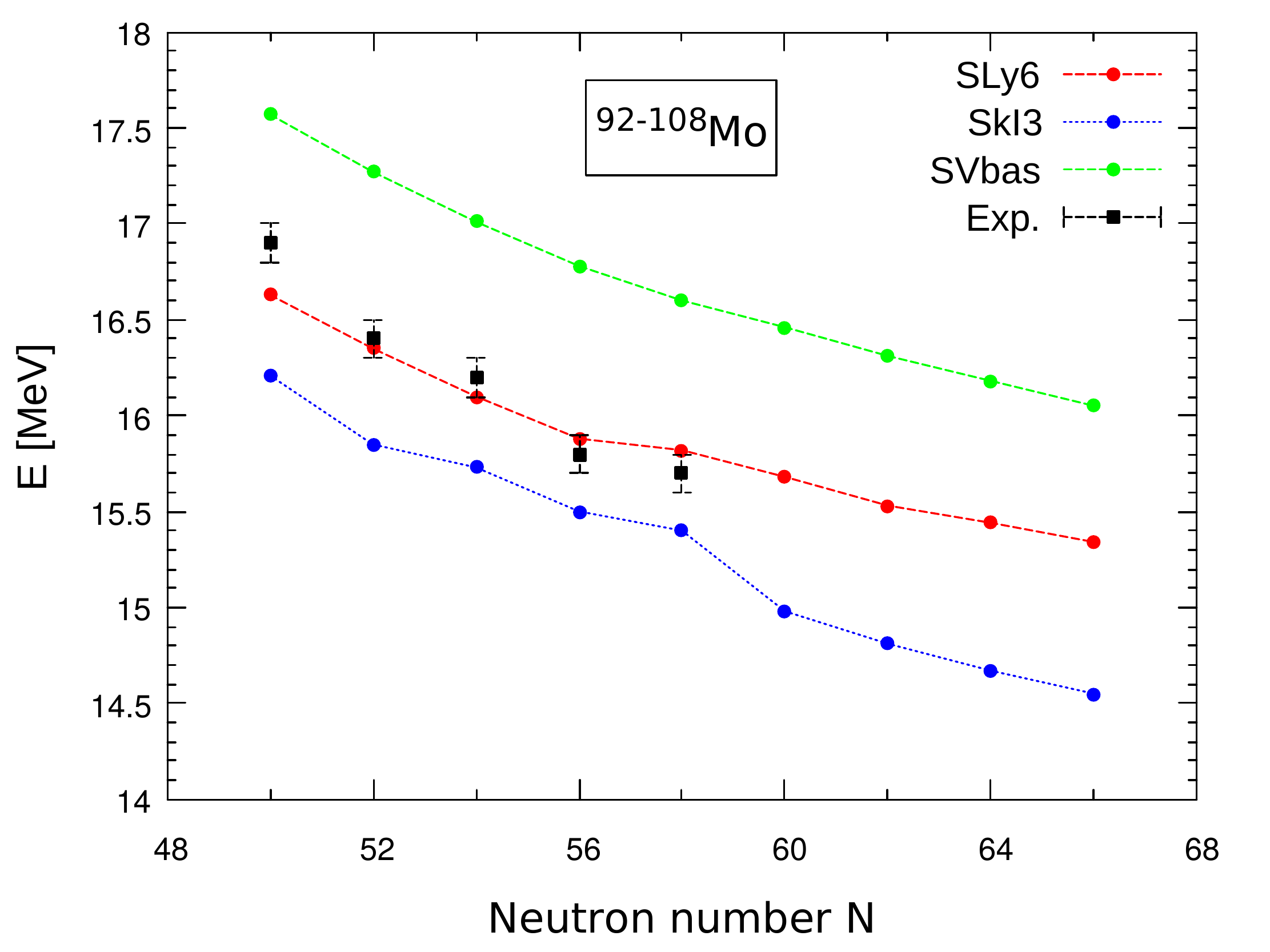} 
	\caption{(Color online) The resonance energies for $^{92-108}\text{Mo}$ nuclei calculated with the Skyrme forces SLy6, SkI3, SVbas, compared with the available experimental data\cite{beil1974}.}
	\label{comp_E}
\end{figure}
Fig.\ref{E&a_s} displays the variation of resonance energy, for $^{92}\text{Mo}$ and $^{100}\text{Mo}$ plotted in Fig.\ref{comp_gdr}, as a function of $\bm{a_{s}}$. It is clearly seen that with decreasing the asymmetry energy $\bm{a_{s}}$ from 34.80 MeV (for SkI3) to 30 MeV (for SVbas), the peak position is moving towards the high energy region. Thus, there is a clear correlation between the energy of the IVGDR and the asymmetry energy $\bm{a_{s}}$. In addition, the value $a_{s} = 32 MeV$ is well consistent with the experimental data to study GDR in deformed nuclei in the framework of TDHF method. It should be taken care for this correlation because our study is restricted only to three Skyrme forces.
We point out that other nuclear matter properties of the Skyrme forces affect the GDR energy, such as the sum rule enhancement factor $\kappa$ \cite{reinhard2009,Bonasera2019} and the isovector effective mass $m_{1}^*/m$ which are related by $m_{1}^*/m=1/(1+\kappa)$ (For more details see Refs.\cite{nesterenko2007,mennana2021}). In Ref. \cite{colo2008}, Colo and al. have found ,for a set of 20 Skyrme interactions, a strong linear correlation between the value of mean energies and the square root of S($\rho$)(1+$\kappa$) at $\rho_{0}$ = 0.1 fm$ ^{-3} $ where S($\rho$) is asymmetry energy and $\kappa$ is the enhancement factor. But they did not find a direct correlation between $ E_{GDR}$ and S($\rho$)$\equiv a_{s}$.

\begin{figure}[!ht]
	\centering
	\includegraphics[width=1.0\textwidth]{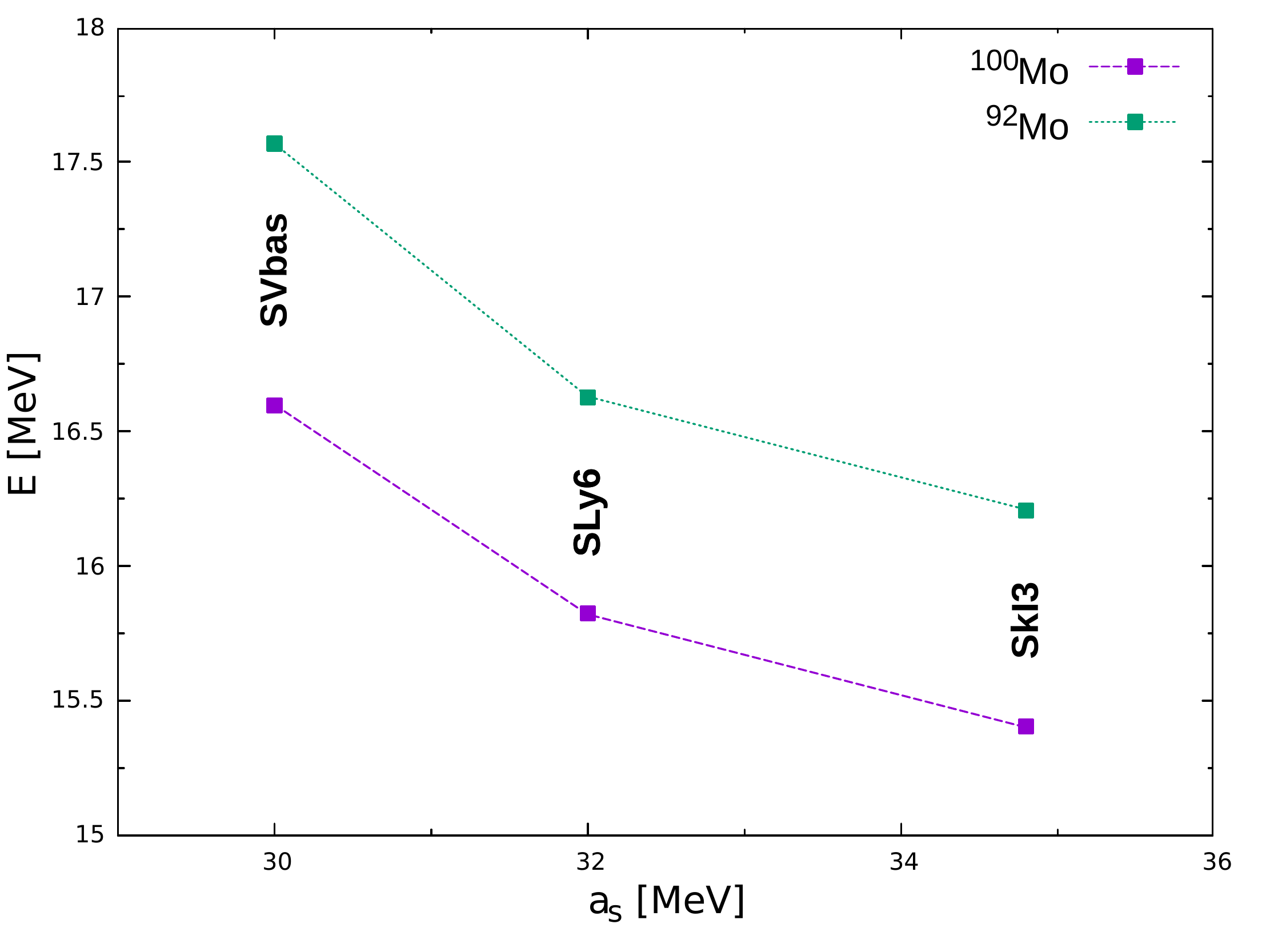} 
	\caption{(Color online) The resonance energy $E$ for $^{92,100}\text{Mo}$ nuclei as a function of the asymmetry energy $a_{s}$.}
	\label{E&a_s}
\end{figure}
\section{Conclusion}\label{sec5}
\qquad The isovector giant dipole resonance IVGDR has been investigated in $^{92-108}\text{Mo}$ nuclei. The investigations have been done within the framework of TDHF method based on the Skyrme functional. All calculations (static and dynamic) were performed with Skyrme force SLy6. We have calculated ground-state properties of these nuclei, especially deformation parameters ($\beta_{2}$, $\gamma$), and compared with available experimental data\cite{beil1974} and theoretical calculations\cite{delaroche2010}. We have found a shape phase transition from spherical $^{92-98}\text{Mo}$ nuclei to axially deformed $^{100-108}\text{Mo}$ ones. For $^{100}\text{Mo}$ isotope, we have predicted a shape coexistence between oblate ($\gamma = 60 ^\circ$) and triaxial ($\gamma = 24 ^\circ$) shape.  

In dynamic calculations, we have calculated some properties of GDR in nuclei under study. The evolution of dipole moment D$_{m}$(t) in time domain showed that oscillation frequencies along the three axes are the same for spherical $^{92}\text{Mo}$ nucleus, $\omega_{x} = \omega_{y} = \omega_{z}$. For deformed nuclei $^{100-108}\text{Mo}$, the oscillation frequency along the major axis ($\omega_{z}$) is higher than that along the minor axis ($\omega_{x} = \omega_{y}$). Also, from the curve of D$_{m}$(t) we have calculated the resonance energies $ E_{i}$ and compared with available experimental results. 

In addition, the  GDR strength calculated is compared with the available experimental data. The results showed that TDHF method can perfectly reproduce the shape of GDR strengths. The calculated resonance energies $ E_{i}$ agree well with the experimental data and reproduce the general trend of decrease with mass number A. Furthermore, the correlation between $\Delta E$ and deformation parameter $\beta_{2}$ is considered. It confirms that the deformation splitting $\Delta E$ is proportional to deformation of nucleus. Finally, we discussed the effect of asymmetry energy $a_{s}$ on GDR strength. Three forces SkI3, SVbas and SLy6 are considered. The results showed that the calculation is well consistent with  experimental data for the value $a_{s} = 32 MeV$ which corresponds to the force SLy6. This confirms that the SLy6 parametrization is the best among these forces, which is characterized by an intermediate value of $a_{s}$.
\section*{Acknowledgments}
Azzeddine Ait Ben Mennana would like to thank P. D. Stevenson from university of Surrey (Uk) and V. O. Nesterenko from University “Dubna”, Moscow (Rus) for helpful discussions. This work was supported by High Energy Physics and Astrophysics Laboratory, faculty of Science Semlalia University Marrakesh, Morocco.
\bibliographystyle{elsarticle-num}
	\bibliography{biblio}

\begin{thebibliography}{10}
\expandafter\ifx\csname url\endcsname\relax
  \def\url#1{\texttt{#1}}\fi
\expandafter\ifx\csname urlprefix\endcsname\relax\def\urlprefix{URL }\fi
\expandafter\ifx\csname href\endcsname\relax
  \def\href#1#2{#2} \def\path#1{#1}\fi

\bibitem{harakeh2001}
M.~N. Harakeh, A.~Woude, Giant Resonances: fundamental high-frequency modes of
  nuclear excitation, Vol. \textbf{24}, Oxford University Press on Demand,
  (2001).

\bibitem{baldwin1947}
G.~C. Baldwin, G.~S. Klaiber, Phys. Rev. \textbf{71} (3 (1947)).
\newblock \href {https://doi.org/10.1103/PhysRev.71.3}
  {\path{doi:10.1103/PhysRev.71.3}}.

\bibitem{goldhaber1948}
M.~Goldhaber, E.~Teller, Phys. Rev. \textbf{74} (1046 (1948)).
\newblock \href {https://doi.org/10.1103/PhysRev.74.1046}
  {\path{doi:10.1103/PhysRev.74.1046}}.

\bibitem{veyssiere1973}
A.~Veyssiere, H.~Beil, R.~Bergere, P.~Carlos, A.~Lepretre, K.~Kernbath, Nuclear
  Physics A \textbf{199} (45 (1973)).

\bibitem{gurevich1976}
G.~Gurevich, L.~Lazareva, V.~Mazur, G.~Solodukhov, B.~Tulupov, Nuclear Physics
  A \textbf{273} (326 (1976)).

\bibitem{carlos1971}
P.~Carlos, H.~Beil, R.~Bergere, A.~Lepretre, A.~Veyssiere, Nuclear Physics A
  \textbf{172} (437 (1971)).
\newblock \href{https://doi.org/10.1016/0375-9474(71)90725-1}{[link]}.
\newline\urlprefix\url{https://doi.org/10.1016/0375-9474(71)90725-1}

\bibitem{carlos1974}
P.~Carlos, H.~Beil, R.~Bergere, A.~Lepretre, A.~De~Miniac, A.~Veyssiere,
  Nuclear Physics A \textbf{225} (171 (1974)).
\newblock \href{https://doi.org/10.1016/0375-9474(74)90373-X}{[link]}.
\newline\urlprefix\url{https://doi.org/10.1016/0375-9474(74)90373-X}

\bibitem{berman1975}
B.~L. Berman, S.~Fultz, Reviews of Modern Physics \textbf{47} (713 (1975)).
\newblock \href {https://doi.org/10.1103/RevModPhys.47.713}
  {\path{doi:10.1103/RevModPhys.47.713}}.

\bibitem{ceruti2017}
S.~Ceruti, F.~Camera, A.~Bracco, A.~Mentana, R.~Avigo, G.~Benzoni, N.~Blasi,
  G.~Bocchi, S.~Bottoni, S.~Brambilla, et~al., Physical Review C \textbf{95}
  (014312 (2017)).

\bibitem{maruhn2005}
J.~A. Maruhn, P.~G. Reinhard, P.~D. Stevenson, J.~R. Stone, M.~R. Strayer,
  Phys. Rev. C \textbf{71} (064328 (2005)).
\newblock \href {https://doi.org/10.1103/PhysRevC.71.064328}
  {\path{doi:10.1103/PhysRevC.71.064328}}.

\bibitem{reinhard2008}
W.~Kleinig, V.~O. Nesterenko, J.~Kvasil, P.-G. Reinhard, P.~Vesely, Phys. Rev.
  C \textbf{78} (044313 (2008)).
\newblock \href {https://doi.org/10.1103/PhysRevC.78.044313}
  {\path{doi:10.1103/PhysRevC.78.044313}}.

\bibitem{colo2008}
L.~Trippa, G.~Colo, E.~Vigezzi, Phys. Rev. C \textbf{77} (061304 (2008)).

\bibitem{yoshida2011}
K.~Yoshida, T.~Nakatsukasa, Phys. Rev. C \textbf{83} (021304 (2011)).
\newblock \href {https://doi.org/10.1103/PhysRevC.83.021304}
  {\path{doi:10.1103/PhysRevC.83.021304}}.

\bibitem{benmenana2020}
A.~A.~B. Mennana, Y.~E. Bassem, M.~Oulne, Physica Scripta \textbf{95} (065301
  (2020)).
\newblock \href{https://doi.org/10.1088/1402-4896/ab73d8}{[link]}.
\newline\urlprefix\url{https://doi.org/10.1088/1402-4896/ab73d8}

\bibitem{beil1974}
H.~Beil, R.~Bergere, P.~Carlos, A.~Lepretre, A.~De~Miniac, A.~Veyssiere, Nuc.
  Phys. A \textbf{227} (427 (1974)).

\bibitem{wang2017}
S.~S. Wang, Y.~G. Ma, X.~G. Cao, W.~B. He, H.~Y. Kong, C.~W. Ma, Phys. Rev. C
  \textbf{95} (054615 (2017)).
\newblock \href {https://doi.org/10.1103/PhysRevC.95.054615}
  {\path{doi:10.1103/PhysRevC.95.054615}}.

\bibitem{ring1980}
P.~Ring, P.~Schuck, The nuclear many-body problem, Springer-Verlag, (1980).

\bibitem{mattiuzzi1997}
M.~Mattiuzzi, A.~Bracco, F.~Camera, W.~Ormand, J.~Gaardh{\o}je, A.~Maj,
  B.~Million, M.~Pignanelli, T.~Tveter, Nuclear Physics A \textbf{612} (262
  (1997)).

\bibitem{pandit2013}
D.~Pandit, B.~Dey, D.~Mondal, S.~Mukhopadhyay, S.~Pal, S.~Bhattacharya, A.~De,
  S.~Banerjee, Physical Review C \textbf{87} (044325 (2013)).

\bibitem{eramzhyan1986}
R.~Eramzhyan, B.~Ishkhanov, I.~Kapitonov, V.~Neudatchin, Physics Reports
  \textbf{136} (229 (1986)).

\bibitem{van1987}
A.~Van~der Woude, Progress in particle and nuclear physics 18 (217 (1987)).

\bibitem{speth1981}
J.~Speth, A.~van~der Woude, Reports on Progress in Physics \textbf{44} (719
  (1981)).

\bibitem{okamoto1958}
K.~Okamoto, Phys. Rev. \textbf{110} (143 (1958)).
\newblock \href {https://doi.org/10.1103/PhysRev.110.143}
  {\path{doi:10.1103/PhysRev.110.143}}.

\bibitem{danos1958}
M.~Danos, Nuclear Physics 5 (23 (1958)).

\bibitem{fuller1958}
E.~G. Fuller, M.~S. Weiss, Phys. Rev. \textbf{112} (560 (1958)).
\newblock \href {https://doi.org/10.1103/PhysRev.112.560}
  {\path{doi:10.1103/PhysRev.112.560}}.

\bibitem{skyrme1956}
T.~H.~R. Skyrme, Phil. Mag. \textbf{1} (1043 (1956)).

\bibitem{mennana2021}
A.~A.~B. Mennana, M.~Oulne, The European Physical Journal Plus \textbf{136} (1
  (2021)).

\bibitem{nesterenko2007}
V.~Nesterenko, W.~Kleinig, J.~Kvasil, P.~Vesely, P.-G. Reinhard, Int. J. Mod.
  Phys. E \textbf{16} (624 (2007)).
\newblock \href {https://doi.org/10.1142/S0218301307006071}
  {\path{doi:10.1142/S0218301307006071}}.

\bibitem{Masur2006}
V.~M. Masur, L.~M. Mel'nikova, Physics of Particles and Nuclei \textbf{37} (923
  (2006)).
\newblock \href {https://doi.org/10.1134/S1063779606060050}
  {\path{doi:10.1134/S1063779606060050}}.

\bibitem{ramakrishnan1996}
E.~Ramakrishnan, T.~Baumann, al., Physical review letters \textbf{76} (2025
  (1996)).
\newblock \href{https://doi.org/10.1103/PhysRevLett.76.2025}{[link]}.
\newline\urlprefix\url{https://doi.org/10.1103/PhysRevLett.76.2025}

\bibitem{gundlach1990}
J.~Gundlach, K.~Snover, J.~Behr, al., Physical review letters \textbf{65} (2523
  (1990)).
\newblock \href{https://doi.org/10.1103/PhysRevLett.65.2523}{[link]}.
\newline\urlprefix\url{https://doi.org/10.1103/PhysRevLett.65.2523}

\bibitem{dirac1930}
P.~A.~M. Dirac, Mathematical Proceedings of the Cambridge Philosophical Society
  \textbf{26} (376 (1930)).
\newblock \href{https://doi.org/10.1017/S0305004100016108}{[link]}.
\newline\urlprefix\url{https://doi.org/10.1017/S0305004100016108}

\bibitem{fracasso2012}
S.~Fracasso, E.~B. Suckling, P.~Stevenson, Physical Review C \textbf{86}
  (044303 (2012)).
\newblock \href{https://doi.org/10.1103/PhysRevC.86.044303}{[link]}.
\newline\urlprefix\url{https://doi.org/10.1103/PhysRevC.86.044303}

\bibitem{sky3d}
B.~Schuetrumpf, P.-G. Reinhard, P.~Stevenson, A.~Umar, J.~Maruhn, Computer
  Physics Communications \textbf{229} (211 (2018)).
\newblock \href {https://doi.org/https://doi.org/10.1016/j.cpc.2018.03.012}
  {\path{doi:https://doi.org/10.1016/j.cpc.2018.03.012}}.

\bibitem{CHABANAT1998}
E.~Chabanat, P.~Bonche, P.~Haensel, J.~Meyer, R.~Schaeffer, Nuclear Physics A
  \textbf{635} (231 (1998)).
\newblock \href {https://doi.org/https://doi.org/10.1016/S0375-94749800180-8}
  {\path{doi:https://doi.org/10.1016/S0375-94749800180-8}}.

\bibitem{rusev2008}
G.~Rusev, R.~Schwengner, F.~D{\"o}nau, M.~Erhard, E.~Grosse, A.~Junghans,
  K.~Kosev, K.~Schilling, A.~Wagner, F.~Be{\v{c}}v{\'a}{\v{r}}, et~al.,
  Physical Review C \textbf{77} (064321 (2008)).

\bibitem{wagner2007}
A.~Wagner, R.~Beyer, M.~Erhard, E.~Grosse, A.~Junghans, J.~Klug, K.~Kosev,
  C.~Nair, N.~Nankov, G.~Rusev, et~al., Journal of Physics G: Nuclear and
  Particle Physics \textbf{35} (014035 (2007)).

\bibitem{ishkhanov2014}
B.~Ishkhanov, I.~Kapitonov, A.~Kuznetsov, V.~Orlin, H.~D. Yoon, Physics of
  Atomic Nuclei \textbf{77} (1362 (2014)).

\bibitem{hartree1928}
D.~R. Hartree, Proc. Camb. Phil. Soc. \textbf{24} (89 (1928)).

\bibitem{bardeen1957}
J.~Bardeen, L.~N. Cooper, J.~R. Schrieffer, Phys. rev. \textbf{108} (1175
  (1957)).

\bibitem{ishkhanov2011}
B.~Ishkhanov, S.~Y. Troshchiev, Moscow University Physics Bulletin \textbf{66}
  (325 (2011)).

\bibitem{ishkhanov2015}
B.~Ishkhanov, I.~Kapitonov, Moscow Univ. Phys. Bulletin \textbf{70} (75
  (2015)).

\bibitem{ishkhanov2021}
B.~S. Ishkhanov, I.~M. Kapitonov, Physics-Uspekhi \textbf{64} (141 (2021)).

\bibitem{maruhn2006}
J.~A. Maruhn, P.-G. Reinhard, P.~D. Stevenson, M.~R. Strayer, Phys. Rev. C
  \textbf{74} (027601 (2006)).
\newblock \href {https://doi.org/10.1103/PhysRevC.74.027601}
  {\path{doi:10.1103/PhysRevC.74.027601}}.

\bibitem{simenel2018}
C.~Simenel, A.~Umar, Progress in Particle and Nuclear Physics \textbf{103} (19
  (2018)).
\newblock \href {https://doi.org/https://doi.org/10.1016/j.ppnp.2018.07.002}
  {\path{doi:https://doi.org/10.1016/j.ppnp.2018.07.002}}.

\bibitem{blocki1979}
J.~B{\l}ocki, H.~Flocard, Physics Letters B \textbf{85} (163 (1979)).
\newblock \href{https://doi.org/10.1016/0370-2693(79)90568-9}{[link]}.
\newline\urlprefix\url{https://doi.org/10.1016/0370-2693(79)90568-9}

\bibitem{stevenson2004}
P.~Stevenson, M.~Strayer, J.~Rikovska~Stone, W.~Newton, International Journal
  of Modern Physics E \textbf{13} ((2004)) 181.
\newblock \href {https://doi.org/10.1142/S0218301304001928}
  {\path{doi:10.1142/S0218301304001928}}.

\bibitem{bonche1976}
P.~Bonche, S.~Koonin, J.~W. Negele, Phys. Rev. C \textbf{13} (1226 (1976)).
\newblock \href {https://doi.org/10.1103/PhysRevC.13.1226}
  {\path{doi:10.1103/PhysRevC.13.1226}}.

\bibitem{kerman1976}
A.~Kerman, S.~Koonin, Annals of Physics \textbf{100} (332 (1976)).
\newblock \href {https://doi.org/https://doi.org/10.1016/0003-49167690065-8}
  {\path{doi:https://doi.org/10.1016/0003-49167690065-8}}.

\bibitem{koonin1977}
S.~E. Koonin, K.~T.~R. Davies, V.~Maruhn-Rezwani, H.~Feldmeier, S.~J. Krieger,
  J.~W. Negele, Phys. Rev. C \textbf{15} (1359 (1977)).
\newblock \href {https://doi.org/10.1103/PhysRevC.15.1359}
  {\path{doi:10.1103/PhysRevC.15.1359}}.

\bibitem{flocard1978}
H.~Flocard, S.~E. Koonin, M.~S. Weiss, Phys. Rev. C \textbf{17} (1682 (1978)).
\newblock \href {https://doi.org/10.1103/PhysRevC.17.1682}
  {\path{doi:10.1103/PhysRevC.17.1682}}.

\bibitem{simenel2012}
C.~Simenel, The European Physical Journal A \textbf{48} (152 (2012)).
\newblock \href{https://doi.org/10.1140/epja/i2012-12152-0}{[link]}.
\newline\urlprefix\url{https://doi.org/10.1140/epja/i2012-12152-0}

\bibitem{nesterenko2006}
V.~Nesterenko, W.~Kleinig, J.~Kvasil, P.~Vesely, P.-G. Reinhard, D.~Dolci,
  Physical Review C \textbf{74} (064306 (2006)).
\newblock \href{https://doi.org/10.1103/PhysRevC.74.064306}{[link]}.
\newline\urlprefix\url{https://doi.org/10.1103/PhysRevC.74.064306}

\bibitem{bender2003}
M.~Bender, P.-H. Heenen, P.-G. Reinhard, Reviews of Modern Physics \textbf{75}
  (121 (2003)).

\bibitem{william1992}
W.~H. Press, S.~A. Teukolsky, W.~T. Vetterling, B.~P. Flannery, Numerical
  Recipes in C : The Art of Scientific Computing 2nd Ed., Cambridge University
  Press, 1992.

\bibitem{reinhard2006}
P.-G. Reinhard, P.~D. Stevenson, D.~Almehed, J.~A. Maruhn, M.~R. Strayer, Phys.
  Rev. E \textbf{73} (036709 (2006)).
\newblock \href {https://doi.org/10.1103/PhysRevE.73.036709}
  {\path{doi:10.1103/PhysRevE.73.036709}}.

\bibitem{takigawa2017}
K.~W. N.~Takigawa, "Fundamentals of Nuclear Physics", Springer Japan, (2017).

\bibitem{raman2001}
S.~RAMAN, C.~NESTOR, P.~TIKKANEN, Atomic Data and Nuclear Data Tables
  \textbf{78} (1 (2001)).
\newblock \href {https://doi.org/https://doi.org/10.1006/adnd.2001.0858}
  {\path{doi:https://doi.org/10.1006/adnd.2001.0858}}.

\bibitem{delaroche2010}
J.-P. Delaroche, M.~Girod, J.~Libert, H.~Goutte, S.~Hilaire, S.~P{\'e}ru,
  N.~Pillet, G.~Bertsch, Physical Review C \textbf{81} (014303 (2010)).
\newblock \href{https://doi.org/10.1103/PhysRevC.81.014303}{[link]}.
\newline\urlprefix\url{https://doi.org/10.1103/PhysRevC.81.014303}

\bibitem{colo2020}
G.~Col{\`o}, D.~Gambacurta, W.~Kleinig, J.~Kvasil, V.~O. Nesterenko,
  A.~Pastore, Phys. Lett. B \textbf{811} (135940 (2020)).

\bibitem{stachel1982}
J.~Stachel, N.~Kaffrell, E.~Grosse, H.~Emling, H.~Folger, R.~Kulessa,
  D.~Schwalm, Nuclear Physics A \textbf{383} (429 (1982)).

\bibitem{wrzosek2012}
K.~Wrzosek-Lipska, Physical Review C \textbf{86} (064305 (2012)).

\bibitem{shi2022}
Y.~Shi, P.~Stevenson, Chinese Phys. C \textbf{47} (034105 (2023)).

\bibitem{garg2018}
U.~Garg, G.~Col{\`o}, Progress in Particle and Nuclear Physics \textbf{101} (55
  (2018)).
\newblock \href{https://doi.org/10.1016/j.ppnp.2018.03.001}{[link]}.
\newline\urlprefix\url{https://doi.org/10.1016/j.ppnp.2018.03.001}

\bibitem{steinwedel1950}
H.~Steinwedel, J.~Jensen, Z.Naturforsch 5A (413 (1950)).

\bibitem{reinhard2009}
P.~Kl\"upfel, P.-G. Reinhard, T.~J. B\"urvenich, J.~A. Maruhn, Phys. Rev. C
  \textbf{79} (034310 (2009)).
\newblock \href {https://doi.org/10.1103/PhysRevC.79.034310}
  {\path{doi:10.1103/PhysRevC.79.034310}}.

\bibitem{REINHARD1995}
P.-G. Reinhard, H.~Flocard, Nuc. Phys. A \textbf{584} (467 (1995)).
\newblock \href {https://doi.org/https://doi.org/10.1016/0375-94749400770-N}
  {\path{doi:https://doi.org/10.1016/0375-94749400770-N}}.

\bibitem{Bonasera2019}
G.~Bonasera, S.~Shlomo, D.~Youngblood, Y.-W. Lui, J.~Button, Nuc. Phys. A
  \textbf{992} (121612 (2019)).

\end{thebibliography}
\end{document}